\documentclass[lettersize,journal]{IEEEtran}
\usepackage{amsmath,amsfonts}
\usepackage{algorithmic}
\usepackage{algorithm}
\usepackage{array}
\usepackage{textcomp}
\usepackage{stfloats}
\usepackage{url}
\usepackage{verbatim}
\usepackage{graphicx}
\usepackage{cite}
\usepackage{soul}
\usepackage{amsthm}
\usepackage{mathtools}
\usepackage{stmaryrd}
\SetSymbolFont{stmry}{bold}{U}{stmry}{m}{n}
\usepackage[dvipsnames]{xcolor}
\usepackage{float,wrapfig,color,comment}
\usepackage[caption=false]{subfig}
\usepackage{bm}
\usepackage{amssymb}
\usepackage{pgfgantt}
\usepackage{multirow}
\usepackage{soul}
\usepackage{bbding}
\usepackage{enumerate}
\usepackage{booktabs}
\usepackage{tabularx}
\usepackage{siunitx}
\usepackage{optidef}

\newtheorem{lemma}{\textbf{Lemma}}
\newtheorem{theorem}{\textbf{Theorem}}

\newtheorem{assumption}{\textbf{Assumption}}
\newcommand{\argmin}{\mathop{\mathrm{argmin}}}

\newcounter{relctr} %% <- counter for relations
\everydisplay\expandafter{\the\everydisplay\setcounter{relctr}{0}} %% <- reset every eq
\newcommand\labelrel[2]{%
  \begingroup
    \refstepcounter{relctr}%
    \stackrel{\textnormal{(\alph{relctr})}}{\mathstrut{#1}}%
    \originallabel{#2}%
  \endgroup
}
\AtBeginDocument{\let\originallabel\label} %

\begin{document}

\title{Online Client Scheduling and Resource Allocation for Efficient Federated Edge Learning}

\author{Zhidong~Gao,~\IEEEmembership{Student~Member,~IEEE,} Zhenxiao~Zhang,~\IEEEmembership{Student~Member,~IEEE,} Yu~Zhang,~\IEEEmembership{Student~Member,~IEEE,} Tongnian~Wang,~\IEEEmembership{Student~Member,~IEEE,} Yanmin~Gong,~\IEEEmembership{Senior~Member,~IEEE,} and Yuanxiong~Guo,~\IEEEmembership{Senior~Member,~IEEE}
\IEEEcompsocitemizethanks{\IEEEcompsocthanksitem Z. Zhang, Z. Gao, Y. Zhang, and Y. Gong are with the Department of Electrical and Computer Engineering, The University of Texas at San Antonio, San Antonio, TX, 78249. T. Wang and Y. Guo are with the Department of Information Systems and Cyber Security, The University of Texas at San Antonio, San Antonio, TX, 78249. E-mail: \{zhidong.gao@my., zhenxiao.zhang@my., yu.zhang@my., tongnian.wang@my., yanmin.gong@, yuanxiong.guo@\}utsa.edu}
}




\maketitle

\begin{abstract}
Federated learning (FL) enables edge devices to collaboratively train a machine learning model without sharing their raw data. Due to its privacy-protecting benefits, FL has been deployed in many real-world applications. However, deploying FL over mobile edge networks with constrained resources such as power, bandwidth, and computation suffers from high training latency and low model accuracy, particularly under data and system heterogeneity. In this paper, we investigate the optimal client scheduling and resource allocation for FL over mobile edge networks under resource constraints and uncertainty to minimize the training latency while maintaining the model accuracy. Specifically, we first analyze the impact of client sampling on model convergence in FL and formulate a stochastic optimization problem that captures the trade-off between the running time and model performance under heterogeneous and uncertain system resources. To solve the formulated problem, we further develop an online control scheme based on Lyapunov-based optimization for client sampling and resource allocation without requiring the knowledge of future dynamics in the FL system. Extensive experimental results demonstrate that the proposed scheme can improve both the training latency and resource efficiency compared with the existing schemes. 
\end{abstract}


\section{Introduction}
\IEEEPARstart{E}{dge} devices, such as mobile phones, smartwatches, and drones, are being increasingly adopted in our daily life and across a wide range of industries. They generate huge amounts of data at the network edge, which need to be processed in a timely fashion to enable real-time decision-making. The traditional machine learning paradigm involves transmitting data over a network to a central cloud server for processing, resulting in significant latency, communication overhead, and privacy issues. Federated learning (FL) is an emerging learning paradigm that enables multiple agents to collaboratively learn from their respective data under the orchestration of a server without sharing their raw data~\cite{mcmahan2017communication}. 

% The standard framework of FL contains an iterative learning process with multiple communication rounds. Each round consists of four stages: (i) a subset of edge devices download the current global model from the server; (ii) each edge device computes a local model based on its local training data in parallel; (iii) edge devices upload their models to the server; (iv) the server aggregates the device models to obtain a new global model. By keeping raw data locally at each edge device, close to where data are generated, FL provides better privacy protection and enjoys higher efficiency compared with the traditional machine learning paradigm. 

Deploying FL at the mobile edge is particularly challenging due to system heterogeneity. Note that edge devices usually have different hardware resources (e.g., CPU frequency, memory, storage capacity, and battery level) and operate under different communication environments. This results in different processing delays for edge devices in the local updating stage. Since the server has to wait for all participating devices to finish the local updating before proceeding to the next round, the total training speed is constrained by the slowest device. 

Meanwhile, FL faces challenges in data heterogeneity where non-IID data distribution across devices hinders the convergence behavior and can even lead to divergence under the highly heterogeneous case \cite{li2020federated}. The challenges become more significant when data heterogeneity and system heterogeneity co-exist. For instance, when a device has little battery energy or a bad network connection, it cannot participate in the training even if its data is vital to training a satisfactory model. In this case, it is helpful for these devices to strategically plan their engagement in the learning process to conserve energy and improve the total learning efficiency.

% marfoq2021federated,jiang2020improving, shen2021agnostic,reisizadeh2020fedpaq,sattler2019robust
Some prior studies have investigated the impact of system heterogeneity aiming to improve communication, computation, and energy efficiency in FL~\cite{rui2021federated, zhang2023communication,9425020,zhang2022scalable}. However, these research works did not consider the potential variability of resources (e.g., battery level and data rate) at edge devices and mainly follow a uniform random sampling process to select edge devices as described in the classic FedAvg algorithm~\cite{mcmahan2017communication}. Hence, the stragglers who possess fewer resources or have a bad communication environment are chosen with the same likelihood as other devices, leading to slow training speed. Another line of research in FL aims to mitigate the impact of data heterogeneity. They either design updating rules based on sample quantities to address the data imbalance issue~\cite{wang2021addressing,yang2021federated} or develop adaptive FL updating algorithms ~\cite{karimireddy2020scaffold,li2021ditto} to address the diverging model update directions caused by different data distributions. However, fewer works have jointly considered the negative impacts of data and system heterogeneity. 

In this paper, we propose a novel Lyapunov-based Resource-efficient Online Algorithm (LROA) for FL over the mobile edges. It strategically selects the participating edge devices at the beginning of each communication round based on their available hardware resources, training data sizes, and communication environments. Meanwhile, computation and communication resources are jointly optimized in each communication round. These enable FL to operate under high degrees of data and system heterogeneity, achieving higher model accuracy, lower training latency, and better resource efficiency. 

Specifically, we analyze the convergence bound of the FL algorithm with arbitrary client sampling under data and system heterogeneity. Based on the convergence bound, we formulate a stochastic optimization problem that captures the trade-off between running time and model convergence under the energy constraint and uncertainty rooting from the random communication environment. The form of the optimization problem is compatible with the Lyapunov \textit{drift-plus-penalty} framework. It enables us to solve the problem online without prior knowledge of system statistics. To demonstrate the effectiveness of our proposed method, we conduct experiments on two benchmark datasets, CIFAR10 and FEMNIST, under various data and system heterogeneity settings. The experimental results illustrate that our method significantly outperforms baselines in terms of the total training latency and convergence speed.

In summary, our main contributions are stated as follows:
\begin{itemize}
    \item We formulate a stochastic optimization problem that captures the trade-off between the running time and model convergence. Then, we design an efficient algorithm based on the Lyapunov optimization and successive upper-bound minimization (SUM) techniques to solve the problem online.
    \item We conduct the convergence analysis for the FL under arbitrary client sampling and derive an error bound that reveals the impact of client sampling under assumptions of non-convex loss function and non-IID data distribution. 
    \item We propose LROA, a novel online resource allocation and client sampling strategy for FL with heterogeneous resources, achieving high accuracy, low latency, and high resource efficiency over mobile edge networks under uncertain environments.
    \item We conduct extensive experiments on several datasets and heterogeneous settings to evaluate the effectiveness of the proposed method. The simulation result shows our method saves up to $50.1\%$ total training latency compared with baselines. 
\end{itemize}

% The remainder of this paper is organized as follows. In Section~\ref{sec:related}, we provide an overview of the related works. Section~\ref{sec:sys_model} describes the system model. We present the convergence analysis in Section~\ref{sec:converge} and formulate the optimization problem in Section~\ref{sec:problem}. Lyapunov-based solution approach is described in Section~\ref{sec:analysis}. We perform extensive experiments to evaluate the proposed algorithm in Section~\ref{sec:exp}. Finally, we conclude this paper in Section~\ref{sec:conclusion}.

% ----------------------------------------
\section{Related works}\label{sec:related}
%---------------------------------------
% wang2021field, lim2020federated, 9141214,
Efficient FL implementation faces two substantial challenges: system and data heterogeneity, both capable of degrading model accuracy and increasing energy consumption and training latency~\cite{zhang2022scalable, guo2022hybrid}. To overcome these challenges,  most of the existing works focus on the resource aspect while ignoring client sampling in FL \cite{yang2020energy, tran2019federated,dinh2020federated}. Recently, a few studies have tried to investigate the impact of client sampling with resource allocation~\cite{nguyen2020efficient, luo2021cost, luo2021costdesign}. However, the sampling probability of each client is assumed to be fixed across the rounds in these works, which cannot capture the dynamic changes in communication and computation resources. On the other hand, there is substantial literature to study the client sampling policy of FL solely from the learning perspective ~\cite{chen2022optimal,rizk2020federated,nguyen2020fast, cho2021client}. The key idea of these studies is to select ``important'' clients with higher probabilities, and the importance is determined either by their local gradient~\cite{chen2022optimal,rizk2020federated,nguyen2020fast} or their local training loss~\cite{cho2021client}. These approaches work well under mild system heterogeneity. However, they may diverge or suffer from high training latency and resource cost under the high degree of system heterogeneity. As shown later in Section \ref{sec:samp}, it is important to take both system heterogeneity and potential dynamics into account when designing the client sampling policy. 

A few recent works mostly related to ours are~\cite{luo2022tackling},~\cite{Peraz2022commun} and~\cite{wang2023federated}, which also consider client sampling from both system and data heterogeneity perspectives. Specifically,  Luo et al.~\cite{luo2022tackling} proposed an adaptive client sampling algorithm to minimize the convergence time under system and data heterogeneity. However, their analysis is restricted to the convex model and ignores the influence of communication and computation on training latency. Perazzone et al.~\cite{Peraz2022commun} proposed a joint client sampling and power allocation scheme that minimizes the
convergence bound and the average communication time under a transmit power constraint. However, they have not considered the resources consumed for local computation. Wang et al.~\cite{wang2023federated} proposed an optimization problem to minimize the convergence bound under the resource constraints by adaptively determining the compressed model update and the probability of local updating. However, their resource model relies on several simplified assumptions (e.g., linear computation cost), which lack a detailed analysis for practical FL systems. Deng, et al.~\cite{deng2022blockchain} utilizes a Lyapunov-based approach to maximize the Long-Term Average (LTA) training data size under LTA energy consumption constraint by solving a mix-integer problem. While they consider a blockchain-assisted FL system, the block mining time and energy consumption are the cores of their formulated optimization problem. Different from these studies, our goal in this paper is to minimize the total learning latency while maintaining the model accuracy by jointly optimizing the client sampling policy, communication, and computation resource allocations under both system and data heterogeneity.

%----------------------------------------
\section{System Modeling}\label{sec:sys_model}
\begin{table}[t]
  \caption{Summary of main notations.}
  \label{tab:notations}
  \centering
  \begin{tabular}{cc}
    \toprule
    Notation & Definition\\
    \midrule
    $n$ &  Index for device\\
    % $k, k^{\prime}$ & Index for device\\
    % $l$ & Index for global round\\
    % $r$ & Index for edge round\\
    % $s$ & Index for local iteration\\
    $t$ & Index for global round\\
    $N$ & Total number of devices\\
    $M$ & Model size in storage\\
    $\bm{\theta}_n^{t}$ &  Local model of device $n$\\ 
    $E$ & Local epoch number \\
    % $\xi_n^i$ & A training sample \\
    % $\ell(\cdot)$ & Loss function \\
    % d & Number of model parameters \\
    % $m$ & Total number of edge servers/clusters\\
    $[N]$ & \{1, 2, \ldots, $N$\}\\
    $K$ & Sampling frequency\\
    $\mathcal{K}^t$ & Set of selected devices in round $t$\\
%    $r$ & Number of devices in $\mathcal{S}^t$\\
    % $n_i$ & Number of devices in cluster $i$\\
    % $\mathcal{G}$ & Communication graph for edge backhaul\\ 
%    $\bm{x}, \bm{x}^{\prime}$ & Models\\
%    {\color{red}$\bm{x}^{(k)}$ & Device $k$'s model}\\
    
    $q_n^t$  & Probability of client $n$ to be selected \\
    ${D}_n$ & Local dataset size of device $n$\\
    % ${D}$ & Total training data size\\
    $w_n$ & Local dataset fraction of device $n$\\
%    $z$ & A data sample\\
    % $F_k(\cdot)$ & Local objective function of device $k$\\
    $F_n(\cdot)$ & Local objective function of device $n$ \\
    
    $N_0$  & Background noise power \\
    $r_{n,d}$  & Download rate of device $n$ \\
    % $B$ & Total communication bandwidth of the server\\
    $B_n$ & Communication bandwidth of device $n$\\
    
    $\bm{h}_n^t$ & Channel gain between device $n$ and the server\\ 
    $c_n$  & CPU cycles per sample \\
    $p_n^t$ & Transmission power of device $n$\\
    $\alpha_n$  & Computing capacitance coefficient \\
    $f_n^t$ & Computing frequency of device $n$\\
    $f_{n}^{\text{min}},f_{n}^{\text{max}}$  & Minimum and maximum CPU frequency \\
    $p_{n}^{\text{min}},p_{n}^{\text{max}}$  & Minimum and maximum communication power \\
    
    $\mathcal{T}_{n,u}^{t,\text{com}}, \mathcal{T}_{n,d}^{t,\text{com}}$ & Uploading and Downloading time \\
    $\mathcal{E}_{n}^{t,\text{cmp}}, \mathcal{E}_{n}^{t,\text{cmp}}$ & Computing and communication energy \\
    $\mathcal{T}_{n}^{t,\text{cmp}}$ & Local computing time \\ 
    $\mathcal{\bar{E}}_n$  & Energy budget of device $n$ \\
    
    $L(t)$ & Quadratic Lyapunov function \\ 
    $Q_n^t$ & Virtual energy consumption queue \\
    $\Delta(t)$ & One-slot conditional Lyapunov drift \\
    % $\Delta_i^t$ & Model updates of device $i$\\
    % $\mathbf{g}_i^{t,s}$ & Stochastic gradient of device $i$\\
    % $\eta$ & Local learning rate\\
%    $\Theta^{(k)}$ & A mini-batch of samples for device $k$\\
    % $\tau$ & Aggregation period\\
    % $C$ & Clipping threshold\\
    
%    $\mathbf{A}^{t}$ & $\randk_k$ projection matrix\\
    % $\mathbf{x}_{i}^{t}$ & Transmit signal of device $i$\\
    % $\pi$ & Number of gossip steps per round\\
    % $\mathbf{z}^t$ & Random Gaussian noise\\
    % $\mathbf{y}^t$ & Received signal by server\\
    
    % $\beta^t$ & Alignment coefficient\\
    % $C_1$ & Bounded gradient coefficient\\
    % $\zeta_i^{2}$ & Bounded variance coefficient\\
   % $\gamma^2, \kappa^2$ & Bounded dissimilarity coefficients\\
   % $\epsilon$, $\delta$ & differential privacy parameters\\
%   $\delta$ & Failure probability\\
%   $p$ & Compression ratio\\
  \bottomrule
\end{tabular}
\end{table}
%----------------------------------------
% In this section, we first introduce the FL system. Then, we present the learning algorithm with adaptive resource control and client sampling. Finally, the detailed communication and computation models are described.

%--------------------------------
\subsection{FL System}\label{sec:fl_basic}
%--------------------------------
% \begin{figure}
%     \centering
%     \includegraphics[width=0.4\textwidth]{./figure/Sys.png}
%     \caption{Workflow of the FL system.}\label{fig:system}
% \end{figure}
We consider an FL system involving $N$ edge devices (or clients) and one server. Each edge device $n \in [N]$ holds a local training dataset $\mathcal{D}_{n}=\{\xi_{n}^{i}\}_{i=1}^{D_n}$ where $\xi_{n}^{i}$ is a training example and $D_n$ is the size of the local training dataset. The total number of training examples across $N$ edge devices is $D = \sum_{n=1}^{N}D_n$. Additionally, we define $\ell(\bm{\theta};\xi_{n}^{i})$ as the per-sample loss function, e.g., cross-entropy or mean square error, where $\bm{\theta} \in \mathbb{R}^{d}$ denotes the model parameters. Then, the local objective of client $n$ can be expressed as
\begin{equation}\label{equ:local}
F_{n}(\bm{\theta}) \coloneqq \frac{1}{D_n} \sum_{i \in \mathcal{D}_n} \ell(\bm{\theta};\xi_{n}^{i}).
\end{equation}
Let $w_{n}= D_n / D $ be the data weight of $n$-th edge device such that $\sum_{n=1}^{N}w_{n}=1, 0 < w_{n} \leq 1$. The goal of the FL is to minimize the following global objective:
\begin{equation}\label{equ:opt}
\min_{\bm{\theta}} F(\bm{\theta}) \coloneqq \sum_{n=1}^{N} w_{n} F_{n}(\bm{\theta}).
\end{equation}

\subsection{FL with Adaptive Resource Control and Client Sampling}\label{sec:samp}
% add more description
The most popular algorithm to solve~\eqref{equ:opt} is FedAvg~\cite{mcmahan2017communication}. Specifically, at the beginning of $t$-th FL round, the server \emph{uniformly} samples a subset of edge devices $\mathcal{K}^{t}$ and broadcasts the global model $\bm{\theta}^{t}$ to the devices. Then, each edge device $n\in\mathcal{K}^{t}$ initializes the local model $\bm{\theta}_n^{t,0}$ by the received global model and runs $E$ epochs SGD based on its local dataset to update the local model. Next, each edge device uploads the local model $\bm{\theta}_n^{t, E}$ to the server, which aggregates the received local models to compute the latest global model $\bm{\theta}^{t+1}$. This process repeats multiple rounds until satisfying the convergence criteria.

One common assumption among FedAvg and its variants~\cite{li2020federated, karimireddy2020scaffold} is that clients are uniformly sampled at each round. However, this assumption does not align with the practical FL systems, as the client may drop out of the training due to various reasons, e.g., network failure or congestion. Moreover, the stragglers are sampled with the same probability, which significantly prolongs the training latency. To improve the efficiency of the FL system, we propose to adopt adaptive sampling so that the stragglers are less likely to be selected during the training.

Building upon recent research \cite{li2019convergence}, we assume that, in round $t$, the server forms a sampled edge device set $\mathcal{K}^{t}$ by sampling $K$ times with replacement from $N$ edge devices. %Note that $\mathcal{K}^{t}$ is a multiset in which an edge device may appear more than once. The aggregation weight of each edge device $n$ is multiplied by the frequency it appears in $\mathcal{K}^{t}$. 
Here, the sampling probability $q_n^t$ of edge device $n$ in round $t$ should satisfy the following constraints:
\begin{equation}\label{equ:cons1}
\sum_{n=1}^{N} q_{n}^{t} = 1, \quad \quad q_{n}^{t} \in(0,1], ~~ \forall n \in [N].
\end{equation}

We summarize the FL algorithm with adaptive client sampling in \textbf{Algorithm~\ref{algorithm-1}}. Specifically, we assume the server to be the coordinator for the FL system. It collects the device-specific parameters from devices, including the CPU cycle per sample $c_n$, the size of the training data $D_n$, energy budget $\mathcal{\bar{E}}_n$, the capacitance coefficient $\alpha _n$, and decision boundaries $f_{n}^{\text{min}},f_{n}^{\text{max}}, p_{n}^{\text{min}}, p_{n}^{\text{max}}$, before the training starts. Other input parameters are determined based on performance criteria, e.g., sampling frequency $K$ and the local epoch $E$, or observable by the server, e.g., background noise power $N_0$ and bandwidth $B$. At the beginning of $t$-th round, each edge device $n$ records its channel gain $h_n^t$ and transmits it to the server (line~\ref{alg1:device_record_upload}). Server determines the transmission power $p_n^t$, CPU frequency $f_n^t$, and sampling probability $q_n^t$ for each client by \textbf{Algorithm~\ref{algorithm-2}} (line~\ref{alg1:server_alg2}), which will be elaborated in Section~\ref{sec:analysis}. Then, the server chooses the set of training devices $\mathcal{K}^t$ by sampling $K$ times according to the probability $\{q_n^t, \forall n\}$ (line~\ref{alg1:server_sample}), and broadcasts the latest global model $\bm{\theta}^{t}$ to all edge devices in $\mathcal{K}^{t}$ (line~\ref{alg1:broadcast}). Each selected edge devices $n\in\mathcal{K}^t$ initializes its local model $\bm{\theta}_{n}^{t,0}$ to the received global model $\bm{\theta}^{t}$ and downloads the control parameters $f_n^t$ and $p_n^t$ (line~\ref{alg1:device_initial}). Next, the selected edge devices perform $E$ epochs of SGD to update their local model $\bm{\theta}_{n}^{t,0}$ (lines~\ref{alg1:local_update}). %Here, local GD consists of two steps: (i) Learner computes the gradient  $\nabla F_n(\bm{\theta}_{n}^{t,s},\mathcal{D}_{n})$ of the local objective function (line 6). (ii) Learner updates the local model with learning rate $\eta$ and gradient $\nabla F_n(\bm{\theta}_{n}^{t,s},\mathcal{D}_{n})$ (line 7). %
After that, all selected edge devices upload their local model updates $(\bm{\theta}_{n}^{t,E}-\bm{\theta}^{t})$ to the server (line~\ref{alg1:device_upload_model}). Finally, the server aggregates the local model updates from $\mathcal{K}^{t}$ to compute the global model $\bm{\theta}^{t+1}$ for the next round (line~\ref{alg1:aggreg}). Since the edge devices are sampled with different probabilities, we re-weight each model update using the corresponding probability and sampling times as follows:
\begin{equation}\label{equ:aggre}
\bm{\theta}^{t+1} \xleftarrow[]{} \bm{\theta}^{t}+\sum_{n\in\mathcal{K}^t}\frac{w_n}{Kq_n^t}(\bm{\theta}_n^{t,E}-\bm{\theta}^{t}).
\end{equation}
% (Here~\eqref{equ:aggre} ensures the aggregated model is unbiased. Note that an edge device can be selected multiple times during $K$ times sampling. In this case, we \blue{inversely} rescale the corresponding aggregation weight by the times it was selected.)
Here~\eqref{equ:aggre} ensures the aggregated model is unbiased towards that with full client participation. We provide the proof in Appendix~\ref{append_agg}. Note it aligns with the aggregation method employed in~\cite{luo2022tackling}, which shares a similar client sampling scheme to ours.

%%%%%%%%%%%%%%%%%%%%%%%%%%%%%%%%%%%%%%%%%%%%%%%%%%%%%%%%%%%
\begin{algorithm}[t]\label{alg_fedavg_arbi_samp}
\textbf{Input:} Sampling frequency $K$, local epoch $E$, background noise power $N_{0}$, bandwidth $B$, model size $M$, download rate $r_{n,d}$, CPU cycle per sample $c_{n}, \forall n$, training data size $D_{n}, \forall n$, capacitance coefficient $\alpha_n, \forall n$, energy budget $\bar{\mathcal{E}}_{n}$, decision boundaries $f_{n}^{\text{min}},f_{n}^{\text{max}}, p_{n}^{\text{min}}, p_{n}^{\text{max}}$.\\
% (\textbf{Control parameters:} $\{p_n^t\}, \{f_n^t\}, \{{q}_{n}^{t}\}$ $\sslash$ determined by \textbf{Algorithm 2}) \\
%({\color{magenta}\textbf{Control parameters:} $\{p_n^t\}, \{f_n^t\}, \{{q}_{n}^{t}\}$, which are determined by \textbf{Algorithm 2}} \red{(simplily put them to input Yu: solved)})\\
\textbf{Output:} Final learned model $\bm{\theta}^{T}$ 

\caption{FL with LROA}\label{algorithm-1}
\begin{algorithmic}[1]
\STATE Initial global model $\bm{\theta}^{0}$;
    \FOR{round $t = 0, \ldots, T - 1$} 
    % \STATE Server samples $K$ times using $\bm{q}^{t}$ with replacement to build candidate device set $\mathcal{K}^{t}$ 
    % \STATE Edge devices  and upload them to the server;\label{alg1:ln1}
    \STATE Each edge device online observes $h_n^t$ and uploads it to the server;\label{alg1:device_record_upload}
    \STATE Server determines $\{p_n^t, \forall n\}, \{f_n^t, \forall n\}, \{{q}_{n}^{t}, \forall n\}$ by \textbf{Algorithm 2}; \label{alg1:server_alg2}
    \STATE Server samples $K$ times by $\{q_{n}^{t}, \forall n\}$ to build  $\mathcal{K}^{t}$;\label{alg1:server_sample}
    \STATE Server broadcasts $\bm{\theta}^{t}$ to edge devices in $\mathcal{K}^{t}$;\label{alg1:broadcast}
    % \STATE Broadcast $\bm{\theta}^{t}$ to all devices in $\mathcal{K}^{t}$
        \FOR{ edge device $n \in \mathcal{K}^{t}$ \textbf{in parallel}}
        \STATE Edge device downloads the control parameters $f_n^t, p_n^t$ from the server and initializes its local model $\bm{\theta}_{n}^{t,0} \xleftarrow[]{} \bm{\theta}^{t}$;\label{alg1:device_initial}
        % \STATE $\bm{\theta}_{n}^{t,0} \xleftarrow[]{} \bm{\theta}^{t}$;\label{alg1:ln6}
        \STATE Edge device updates $\bm{\theta}_{n}^{t,0}$ with $E$ epochs to compute $\bm{\theta}_n^{t,E}$ under computation frequency $f_n^t$;\label{alg1:local_update}
        \STATE Edge device uploads $\bm{\theta}_n^{t,E} -\bm{\theta}_{n}^{t,0}$ to the server with communication power $p_n^t$;\label{alg1:device_upload_model}
        % \STATE $\bm{\theta}_{n}^{t,0} \xleftarrow[]{} \bm{\theta}^{t}$
        % % \STATE device LocalUpdate with E epochs.
        %     \FOR{ $s \xleftarrow[]{} 0,1,\ldots, (E-1)$} %\frac{D_n}{B}$ }
        %     \WHILE {true}
        %     \STATE Sample $\{\xi_{n}^{i}\}$ from $\mathcal{D}_{n}$ without replacement.
        %     \STATE Compute mini-batch gradient $\nabla F_{n}(\bm{\theta}_{n}^{t,s}, \{\xi_{n}^{i}\})$. 
        %     \STATE $\bm{\theta}_{n}^{t,s} \xleftarrow[]{} \bm{\theta}_{n}^{t,s} - \eta \nabla F_{n}(\bm{\theta}_{n}^{t,s}, \{\xi_{n}^{i}\})$
        %     \ENDWHILE
        %     \STATE $\bm{\theta}_{n}^{t,s+1} \xleftarrow[]{} \bm{\theta}_{n}^{t,s}$
        %     \ENDFOR
        % % \STATE $\bm{\theta}_{n}^{t+1} \xleftarrow[]{} \bm{\theta}_{n}^{t,E}$
        % % \STATE device uploads local updated model $\bm{\theta}_{n}^{t,E} - \bm{\theta}^{t}$ to server.
        \ENDFOR
    \STATE $\bm{\theta}^{t+1} \xleftarrow[]{} \bm{\theta}^{t}+\sum_{n\in\mathcal{K}^t}\frac{w_n}{Kq_n^t}(\bm{\theta}_n^{t,E}-\bm{\theta}^{t})$;\label{alg1:aggreg}
    \ENDFOR
\end{algorithmic}
\end{algorithm}
%--------------------------------
\subsection{Communication Time}\label{sec:com}
%--------------------------------

Without loss of generality, we assume the frequency-division multiple access (FDMA) protocol is adopted, and the server evenly allocates its communication bandwidth among all selected edge devices. Let $B$ be the total communication bandwidth of the server. Then, the allocated bandwidth for a selected edge device is $B_{n} = {B}/{K}, \forall n \in \mathcal{K}^t$,  where $K$ is the number of selected edge devices at each communication round. Furthermore, we model the channel gain from edge device $n$ to the server as a discrete-time random process $h_{n}^{t}$. According to Shannon's capacity theorem, the achievable uploading transmission rate $r_{n,u}^{t}$ of edge device $n$ at round $t$ is given by
\begin{equation}\label{equ:com_rate}
\begin{split}
r_{n,u}^{t} = B_{n}\log_{2}(1+\frac{h_{n}^{t}p_{n}^{t}}{N_{0}}), \quad \forall n \in \mathcal{K}^t,
% \\
% r_{n,d} = B_{d}\log_{2}(1+\frac{h_{n,d}^{t}P_{0}}{N_{0}}), 
\end{split}
\end{equation}
where $p_{n}^t$ is the edge device's transmission power, $N_{0}$ is the additive Gaussian white noise.

% (Note $h_{n}^{t}$ depends on the communication environment, e.g., atmospheric absorption, the distance from the edge device to the server, and the obstacles (e.g., buildings and terrain) in the communication path, which is random and unknown before online observation.)

Consequently, the communication time for uploading $\mathcal{T}_{n,u}^{t,\text{com}}$ and downloading $\mathcal{T}_{n,d}^{t,\text{com}}$ are 
% \red{(Need to consider both uplink and downlink in the model)}
\begin{align}
\mathcal{T}_{n,u}^{t,\text{com}} &=\frac{M}{r_{n,u}^{t}}= \frac{M K}{B \log_{2}(1+\frac{h_{n}^{t}p_{n}^{t}}{N_{0}})}, \quad \forall n \in \mathcal{K}^t,\label{equ:com_time} \\
\mathcal{T}_{n,d}^{t,\text{com}} &= \frac{M}{r_{n,d}}, \quad \forall n \in \mathcal{K}^t,\label{equ:com_time_d} 
% =\frac{M}{B_{d} \log_{2}(1+\frac{h_{n,d}^{t}P_{0}}{N_{0}})}, 
\end{align}
where $M$ denotes the size of the model update, and $r_{n,d}$ is downloading transmission rate. 

%--------------------------------
\subsection{Computation Time}\label{sec:cmp}
%--------------------------------
Let $c_{n}$ denote the number of CPU cycles required to process a single data sample for edge device $n$. Then, the total number of CPU cycles required to train the local model at each round is $Ec_{n}D_{n}$. Thus, the per-round local computation time for edge device $n$ can be estimated as

\begin{equation}\label{equ:cmp_time}
\mathcal{T}_{n}^{t,\text{cmp}} =  \frac{Ec_{n}D_{n}}{f_{n}^{t}}, \quad \forall n \in \mathcal{K}^t,
\end{equation}
where $f_{n}^{t}$ is the CPU frequency that edge device $n$ used for local model updating at round $t$.

%--------------------------------
\subsection{Per-round Training Time}\label{sec:time}
%--------------------------------
The per-round training time for a client is composed of the model downloading time, local training time, and model update uploading time. Let $\mathcal{T}_{n}^{t}$ denote the per-round time consumption of edge device $n$, then we have 
\begin{equation}\label{equ:time_sum}
\mathcal{T}_{n}^{t} = \mathcal{T}_{n}^{t,\text{cmp}}+\mathcal{T}_{n,u}^{t,\text{com}}+\mathcal{T}_{n,d}^{t,\text{com}}, \quad \forall n \in \mathcal{K}^t.
\end{equation}

% We consider synchronized FL as it can be easily integrated with other FL extensions such as DP-FedAvg~\cite{andrew2021differentially,hu2020personalized} and personalized FL~\cite{karimireddy2020scaffold,li2020federated}. 

In synchronous FL, the server computes the global model after all clients have uploaded their local model. Therefore, the slowest edge device determines the training latency of one global round
\begin{equation}\label{equ:time}
\mathcal{T}^{t} = \underset{n \in \mathcal{K}^{t}}{\max}\{\mathcal{T}_{n}^{t} \},
\end{equation}
where $\mathcal{T}^{t}$ is the wall-clock training time of FL at round $t$.Note that directly optimizing the selected client set $\mathcal{K}^t$ is not feasible since the relationship between sample probabilities $q_{n}^{t}$ and the selected client set $\mathcal{K}^t$ is complex and hard to handle. Therefore, we adopt a similar strategy as in~\cite{luo2022tackling} and approximate the per-round completion time as
\begin{equation}\label{equ:time_appro}
\underset{n \in \mathcal{K}^{t}}{\max}\{\mathcal{T}_{n}^{t} \} \approx \sum_{n=1}^{N} q_{n}^{t}\mathcal{T}_{n}^{t}.
\end{equation} 

%--------------------------------
\subsection{Energy Consumption}\label{sec:eng}
%--------------------------------

In each communication round, the selected edge devices perform $E$ epochs of local training. Following~\cite{yang2020energy}, the computation energy of device $n$ at round $t$ is
\begin{equation}\label{eq_cmp_energy}
    \mathcal{E}_{n}^{t,\text{cmp}} = \frac{E\alpha_{n}c_{n}D_n(f_{n}^{t})^{2}}{2}, \quad \forall n \in \mathcal{K}^t,
\end{equation}
where $\alpha_{n}/2$ is the effective capacitance coefficient of device $n$ (determined by chip architecture).

In practice, the available CPU frequency $f_{n}^{t}$ of an edge device has an upper bound due to hardware limitations. Therefore, we have the following constraint on the CPU frequency:
\begin{equation}\label{equ:cons3}
f_{n}^{\text{min}} \leq f_{n}^{t} \leq f_{n}^{\text{max}}, \quad \forall n \in \mathcal{K}^t,
\end{equation}
where $f_{n}^{\text{max}}$ and $f_{n}^{\text{min}}$ denote the maximum and minimum available CPU frequency of edge device $n$, respectively.

The communication energy of uploading for edge device $n$ at round $t$ can be estimated as
\begin{equation}\label{eq_com_energy}
\mathcal{E}_{n}^{t,\text{com}} = p_{n}^{t} \mathcal{T}_{n,u}^{t,\text{com}} = \frac{p_{n}^{t} MK }{ B \log_{2}(1 + \frac{h_{n}^{t}p_{n}^{t}}{N_{0}})}, \quad \forall n \in \mathcal{K}^t.
% \mathcal{E}_{n,d}^{t,\text{com}} &= p_{n,d}^{t} \mathcal{T}_{n,d}^{t,\text{com}} = \frac{p_{n,d}^{t} MK }{B_{d} }.
\end{equation}
The energy of global model downloading is assumed to be neglectable since the receiving power of edge device is very small compared with its transmission power. Consequently, the total energy consumption of edge device $n$ at round $t$ is
\begin{equation}\label{eq_energy}
 \mathcal{E}_{n}^{t}  = \mathcal{E}_{n}^{t,\text{cmp}} +\mathcal{E}_{n}^{t,\text{com}}, \quad \forall n \in \mathcal{K}^t. 
\end{equation}
Since edge devices, such as mobile phones and smartwatches, are usually power-limited, and the energy consumption of edge device $n$ depends on the random channel gain $h_{n}^{t}$, we consider the following energy constraint on the time-average expected energy consumption of each edge device $n$:
 \begin{equation}\label{equ:cons4}
\lim_{T \to \infty}\frac{1}{T}\sum_{t=0}^{T-1}\mathbb{E}_{h_{n}^{t}}\left[(1-(1-q_{n}^{t})^{K}) \mathcal{E}_{n}^{t}\right] \leq \bar{\mathcal{E}}_{n}, \quad \forall n \in [N],
\end{equation}
Here $(1-(1-q_{n}^{t})^{K})$ is the likelihood of device $n$ being chosen at round $t$. Specifically, for each sampling event, the probability of a client not being selected is $1- q_{n}^{t}$. After $K$ sampling events, the compounded probability of client $n$ not being selected in any of these events is $(1- q_{n}^{t})^{K}$. Then, the probability of a client being selected at least once during $K$ times sampling is $1- (1- q_{n}^{t})^{K}$.

In practice, due to the edge device antenna limitation, we have the following constraint on the transmission power:
\begin{equation}\label{equ:cons2}
\begin{split}
p_{n}^{\min} \leq p_{n}^{t} \leq p_{n}^{\max},  \quad \forall n \in \mathcal{K}^t.
\end{split}
\end{equation}
where $p_{n}^{\min}, p_{n}^{\max}$ denote the minimum and maximum transmission power of edge device $n$.

%-----------------------------------------------------
\section{Convergence Analysis}\label{sec:converge}
%-----------------------------------------------------
In this section, we derive the convergence properties of Algorithm~\ref{algorithm-1} under non-convex and non-IID settings. Before stating our convergence results, we make the following assumptions:

\begin{assumption}[Smoothness]\label{ass:smoothness}
Each local objective function $F_n:\mathbb{R}^d\rightarrow\mathbb{R}$ is $\beta$-smooth for all $n\in [N]$, i.e.,
\[
    \|\nabla F_n(\bm{\theta}) - \nabla F_n(\bm{\theta}^\prime)\|  \leq \beta \|\bm{\theta} - \bm{\theta}^{\prime}\|,    \; \forall \bm{\theta},\bm{\theta}^{\prime}\in \mathbb{R}^d. 
\]
% \[
%     \| g_n(\bm{\theta}) -  g_n(\bm{\theta}^\prime)\|  \leq \beta \|\bm{\theta} - \bm{\theta}^{\prime}\|,    \; \forall \bm{\theta},\bm{\theta}^{\prime}\in \mathbb{R}^d. 
% \]
\end{assumption}

\begin{assumption}[Bounded gradients]\label{ass:bounded_sgd}
There exists a constant $G\geq0$ such that
\[
\|\nabla F_n(\bm{\theta})\|_2^2 \leq G^2, \forall \bm{\theta} \in \mathbb{R}^d, n \in [N].
\]
\end{assumption}

\begin{assumption}[Bounded Dissimilarity]\label{ass:bound_dissimi}
There exist constants $\gamma^2\geq  1, \kappa^2 \geq 0$ such that $\sum_{n=1}^{N}w_n\|\nabla F_n(\bm{\theta})\|^2\leq \gamma^2\|\sum_{n=1}^{N}w_n\nabla F_n(\bm{\theta})\|^2+\kappa^2$. If the data distributions across all devices are IID, then $\gamma^2=1$ and $\kappa^2=0$.
\end{assumption}

Assumptions \ref{ass:smoothness} and \ref{ass:bounded_sgd} are commonly used in the FL literature, e.g.,~\cite{yu2019parallel},~\cite{stich2019local} and~\cite{chen2022optimal}. Assumption \ref{ass:bound_dissimi} captures the dissimilarities of local objective functions under non-IID data distribution, such as~\cite{li2020convergence} and~\cite{ward2020adagrad}. Note that some recent works~\cite{luo2022tackling},~\cite{Peraz2022commun} also consider arbitrary device probabilities for FL and provide convergence analyses, but they only consider convex local objectives and IID data distribution. However, deep neural networks are usually non-convex, and the data are generally non-IID over devices in FL. Thus, our analysis is more general than prior works.

%--------------------------------
% \subsection{Convergence Result}
%--------------------------------
We now state the convergence analysis result in the following theorem.
%%%%%%%%%%%%%%%%%%%%%%%%%%%%%%%%%%%%%%%
\begin{theorem}[Convergence Result with Adaptive Sampling Probabilities]\label{th:convergence}
Under Assumptions \ref{ass:smoothness}, \ref{ass:bounded_sgd} and \ref{ass:bound_dissimi}, if the local learning rate $\eta \leq \min\{{1}/({32E^2 \beta^2\gamma^2}), {1}/({2\sqrt{2}E \beta})\}$, then Algorithm~\ref{algorithm-1} satisfies
\begin{equation}\label{theo_upp}
    \begin{split}
        \frac{1}{T}\sum_{t=0}^{T-1}\mathbb{E}\|\nabla F(\bm{\theta}^t)\|^2\leq  &\frac{4(F(\bm{\theta}^{0})-F^*)}{\eta TE}+8\eta^2\beta^2E^2 \kappa^2\\ 
        &+ \frac{2\beta\eta E G^2}{KT}\sum_{t=0}^{T-1}\sum_{n=1}^{N}\frac{w_n^2}{q_n^t}.
    \end{split}
\end{equation}
% \begin{multline}\label{theo_upp}
% \frac{1}{T}\sum_{t=0}^{T-1}\mathbb{E}\|\nabla F(\bm{\theta}^t)\|^2\leq  \frac{4(F(\bm{\theta}^{0})-F^*)}{\eta TE}+8\eta^2\beta^2E^2 \kappa^2\\ + \frac{2\beta\eta E G^2}{KT}\sum_{t=0}^{T-1}\sum_{n=1}^{N}\frac{w_n^2}{q_n^t}. 
% \end{multline}
\end{theorem}

\begin{IEEEproof} 
The detailed proof is given in Appendix~\ref{append_conv}.
\end{IEEEproof}
%%%%%%%%%%%%%%%%%%%%%%%%%%%%%%%%%%%%%%%
We can see that the convergence bound~\eqref{theo_upp} contains three terms. The first two terms match the optimization error bound in standard FedAvg~\cite{bottou2018optimization}. The third term represents the additional sampling error due to the device sampling. As $K$ increases, the sampling error becomes smaller but at the cost of a potentially large energy consumption and training latency.  
% \begin{corollary}\label{coro}
% Under Assumptions \ref{ass:smoothness}, \ref{ass:bounded_sgd} and \ref{ass:bound_dissimi},  if the learning rate is $\eta = \sqrt{\frac{1}{TE}}$ {\color{red}{when $TE>\max\{(32E^2\beta^2\gamma^2)^2, 8E^2\beta^2\}$},} then 
% \begin{gather*}
%     \frac{1}{T}\sum_{t=0}^{T-1}\mathbb{E}\|\nabla F(\bm{\theta}^t)\|^2  \leq O(\frac{1}{\sqrt{TE}})+O(\frac{\sqrt{E}}{T^{3/2}}\sum_{t=0}^{T-1}\sum_{n=1}^{N}\frac{w_n^2}{q_n^t})
% \end{gather*}
% \end{corollary}

%-----------------------------------------------------
\section{Problem Formulation}\label{sec:problem}
%-----------------------------------------------------
In this paper, we are interested in minimizing the time-average expected training latency while maintaining the model accuracy over a large time horizon. Therefore, the control problem can be stated as follows: for the dynamic FL system, design a control strategy which, given the past and the present random channel gain, chooses the CPU frequency $\{f_{n}^{t}\}$, transmission power $\{p_{n}^{t}\}$ and sampling probability $\{q_{n}^{t}\}$ of edge devices such that the time-average expected training latency is minimized while keeping the high accuracy. It can be formulated as the following stochastic optimization problem:
\begin{align*}
    \textbf{P1:} \quad \min_{\{f_{n}^{t}\},\{p_{n}^{t}\}, \{q_{n}^{t}\}} & \quad \underset{T \xrightarrow{} \infty}{\text{lim}} \frac{1}{T} \sum_{t=0}^{T-1}\sum_{n=1}^{N}\mathbb{E}_{h_{n}^{t}} \left[q_{n}^{t}\mathcal{T}_{n}^{t} +\lambda \cdot \frac{w_n^2}{ q_{n}^{t}}  \right]  \\
    \text{s.t.}  & \quad   (\ref{equ:cons1}), (\ref{equ:com_time_d}), (\ref{equ:com_time}),(\ref{equ:cmp_time}),(\ref{eq_cmp_energy}),(\ref{equ:cons3}),(\ref{eq_com_energy}),  \\
    & \quad (\ref{eq_energy}),(\ref{equ:cons4}),(\ref{equ:cons2}). 
\end{align*}
One challenge of solving this optimization problem lies in the uncertainty of channel state information, which makes Problem $\textbf{P1}$ stochastic. Another challenge is the energy constraint \eqref{equ:cons4} brings the ``time-coupling property" to Problem $\textbf{P1}$. In other words, the current control action may impact future control actions, making the Problem $\textbf{P1}$ more challenging to solve. Moreover, the optimization problem is highly non-convex, which is hard to solve.

% %%%%%%%%%%%%%%%%%%%%%%%%

%-----------------------------------------------------
\section{Resource-Efficient online control policy} \label{sec:analysis}
%-----------------------------------------------------

In this section, we develop an online control algorithm to solve the stochastic optimization problem~\textbf{P1}. Our approach utilizes the Lyapunov optimization framework~\cite{neely2010stochastic}, which eliminates the need for prior knowledge of the FL system and can be easily implemented in real-time.
%-----------------------------------------------------
\subsection{The Lyapunov-Based Approach}\label{sec:lya-app}
%-----------------------------------------------------
The basic idea of the Lyapunov optimization technique is to utilize the stability of the queue to ensure that the time-average constraint is satisfied \cite{neely2010stochastic}. Following this idea, we first construct a \textit{virtual} energy consumption queue $Q_n^t$ for each edge device $n \in [N]$, which represents its backlog of energy consumption at the current round $t$. The updating equation of queue $Q_n^t$ is given by

\begin{equation}
Q_n^{t+1} \coloneqq \max\{Q_n^{t} + a_n^t , 0\},\label{eq_virt_queue}
\end{equation}
where 
\begin{equation}\label{equ:ant}
a_n^t  \coloneqq (1-(1-q_{n}^{t})^{K}) \mathcal{E}_{n}^{t} - \bar{\mathcal{E}}_{n}.
\end{equation}
We can easily show that the stability of the virtual queues \eqref{eq_virt_queue} implies the satisfaction of the energy constraint~\eqref{equ:cons4}. Moreover, we define the quadratic Lyapunov function as follows:
\begin{equation}
L(t)\coloneqq \frac{1}{2}\sum_{n=1}^{N}(Q_n^t)^2.
\end{equation}
For ease of presentation, we use $\bm{Q}^t$ to denote the queue status $\{Q_n^t, \forall n\}$ at round $t$. The one-slot conditional Lyapunov drift can be formulated as
\begin{equation}
\Delta(t)\coloneqq\mathbb{E}\{L(t+1)-L(t)|\bm{Q}^t\},
\end{equation}
where the expectation is taken w.r.t. the randomness of channel gains $h_{n}^t$. Following the Lyapunov optimization framework, we have the Lyapunov \textit{drift-plus-penalty} term
\begin{equation}
\Delta_{V}(t)\coloneqq\Delta(t)+V\mathbb{E}\left\{\sum_{n=1}^{N} \left(q_{n}^{t}\mathcal{T}_{n}^{t}+\lambda \cdot  \frac{w_n^2}{ q_{n}^{t}}\right)|\bm{Q}^t\right\},
\end{equation}
where $V > 0$ is a parameter that controls the trade-off between queue stability and optimality of the objective function. 

We assume the random channel gain $h_n^t$ is IID distributed across different time slots $t$. Then, we have the following lemma for the upper bound of Lyapunov \textit{drift-plus-penalty} term: 

%%%%%%%%%%%%%%%%%%%%%%%%%%%%%%%%%%%%%%%
\begin{lemma}\label{lemma_drift_plus_penalty}
For any feasible action under constraints~\eqref{equ:cons3},~\eqref{equ:cons2},~\eqref{equ:cons1},~\eqref{equ:cons4}, we have
\begin{multline}\label{in_drift_plus_penalty_upper_bound}
\Delta_{V}(t)\leq C+V\mathbb{E}[\sum_{n=1}^{N} \left(q_{n}^{t}\mathcal{T}_{n}^{t}+\lambda \cdot  \frac{w_n^2}{ q_{n}^{t}}\right)|\bm{Q}^t]\\
+\sum_{n=1}^{N} Q_n^t \mathbb{E}[a_n^t|\bm{Q}^t],
\end{multline}
where $C$ is a constant given by
\begin{align*}\label{def_C}
    C\coloneqq \sum_{n=1}^{N}\left\{\left( \bar{\mathcal{T}}_n^{t,\textup{com}} p_{n}^{\max} + \frac{E\alpha_{n}c_{n}D_n(f_{n}^{\max})^{2}}{2}\right)^2 + (\bar{\mathcal{E}}_{n})^2\right\},
\end{align*}
where $\bar{\mathcal{T}}_n^{t,\textup{com}}$ is the upper bound of $\mathcal{T}_{n,u}^{t,\textup{com}}$.
\end{lemma}
%%%%%%%%%%%%%%%%%%%%%%%%%%%%%%%%%%%%%%%%
\begin{IEEEproof}
The proof is provided in Appendix~\ref{append_lemma1}.
\end{IEEEproof}

% Our goal is to minimize the R.H.S of~\eqref{in_drift_plus_penalty_upper_bound} in each time slot $t$ by adequately choosing the control action. 

Now, we present the proposed algorithm. Our goal is to minimize the R.H.S of~\eqref{in_drift_plus_penalty_upper_bound} at time slot $t$ by employing a greedy algorithm.
% The main idea is employing a greedy algorithm to minimize the upper bound given by the right-hand side of~\eqref{in_drift_plus_penalty_upper_bound} in each time slot $t$.

\textit{Lyapunov-Based Resource-Efficient Online Algorithm}: Initialize $\bm{Q}^0=\bm{0}$. At each time slot $t$, observe $h_{n}^t$, and do:
\begin{enumerate}
\item Choose control decisions  $\bm{f}^t=\{f_n^t,\forall n\}$, $\bm{p}^t=\{p_{n}^t,\forall n\}$, and $\bm{q}^t = \{q_n^t,\forall n\}$ for device $n$ as the optimal solution to the following optimization problem:
\begin{align*}
 \textbf{P2:} \quad \min_{\bm{f}^t, \bm{p}^t, \bm{q}^t}\quad & V\sum_{n=1}^{N} \left(q_{n}^{t}\mathcal{T}_{n}^{t}+\lambda \frac{w_n^2}{ q_{n}^{t}}\right)+\sum_{n=1}^{N}Q_n^ta_n(t) \\
 \text{s.t.} \quad &  (\ref{equ:cons1}),(\ref{equ:com_time}),(\ref{equ:cmp_time}),(\ref{equ:time_sum}),(\ref{eq_cmp_energy}),(\ref{equ:cons3}),(\ref{eq_com_energy}), \\
    &(\ref{eq_energy}),(\ref{equ:cons4}),(\ref{equ:cons2}). 
\end{align*}
\item Update $\bm{Q}^t$ regarding to the dynamics~\eqref{eq_virt_queue} and~\eqref{equ:ant}.
\end{enumerate}

Note that Problem \textbf{P2} is non-convex, which is hard to solve. In the following subsection, we develop an efficient algorithm to solve  Problem \textbf{P2}.

%-----------------------------------------------------
\subsection{Efficient Solution Algorithm to \textbf{\textup{P2}}}
%-----------------------------------------------------
By decoupling the optimization of decision variables, we design an alternating minimization-based algorithm to obtain an effective solution to Problem P2. Specifically, we first optimize $\bm{f}^t$ and $\bm{p}^t$ while keeping $\bm{q}^t$ fixed. Then, we update $\bm{q}^t$ utilizing the optimal values of $\bm{f}^t$ and $\bm{p}^t$ obtained in the preceding step. This iterative procedure continues until convergence is achieved.
% The details are outlined as follows.
% In this subsection, we develop an efficient solution algorithm to solve \textbf{P3} based on alternating minimization: we first optimize $\bm{f}^t$ and $\bm{p}^t$ with fixed $\bm{q}^t$, and then update $\bm{q}^t$ based on the optimal $\bm{f}^t$ and $\bm{p}^t$ obtained in the previous step, and the same procedure iterates until convergence. 
% Details are the following:
\subsubsection{Optimization w.r.t. $\bm{f}^t$ and $\bm{p}^t$}
Under a fixed $\bm{q}^t$, we can rewrite the Problem $\textbf{P2}$ as
\begin{align*}
\textbf{P2.1:} \quad \min_{\bm{f}^t, \bm{p}^t} \quad &\sum_{n=1}^{N}Q_n^t\big(1-(1-q_{n}^{t})^{K}) \frac{E\alpha_{n}c_{n}D_n(f_{n}^{t})^{2}}{2}  \\
&+\sum\limits_{n=1}^{N} \frac{MK\big(Q_n^t (1-(1-q_{n}^{t})^{K}) p_{n}^{t} + Vq_{n}^{t}\big)}{B\log(1+\frac{h_{n}^{t}p_{n}^{t}}{N_{0}})} \\
& + V\sum_{n=1}^{N}q_{n}^{t}\frac{Ec_{n}D_n}{f_{n}^{t}} \\
\text{s.t.}  \quad & f_{n}^{\min} \leq f_{n}^{t} \leq f_{n}^{\max},\quad \forall n\in[N] \\
  % & & & p_{n}^{\min} \leq p_{n,u}^{t} \leq p_{n}^{\max},\quad \forall n\in[N]\notag\\
  &  p_{n}^{\min} \leq p_{n}^{t} \leq p_{n}^{\max},\quad \forall n\in[N].
\end{align*}
In Problem \textbf{P2.1}, the decision variables $\bm{f}^t$ and $\bm{p}^t$ are decoupled in both the objective function and constraints. This allows us to solve them separately. Specifically, we have the following two sub-problems:
\begin{align*}
\textbf{P2.1.1:} \quad \min_{\bm{f}^t} \quad &\sum_{n=1}^{N}Q_n^t \big(1-(1-q_{n}^{t})^{K})\frac{E\alpha_{n}c_{n}D_n(f_{n}^{t})^{2}}{2}\\
&+ V\sum_{n=1}^{N}q_{n}^{t}\frac{Ec_{n}D_n}{f_{n}^{t}}\\
\text{s.t.} \quad &f_{n}^{\min} \leq f_{n}^{t} \leq f_{n}^{\max},\quad \forall n\in[N]. 
\end{align*}
\begin{align*}
\textbf{P2.1.2:} \quad \min_{\bm{p}^t} \quad &
\sum_{n=1}^{N} \frac{MK( Vq_{n}^{t}+ Q_n^t (1-(1-q_{n}^{t})^{K}) p_{n}^{t} )}{B\log(1+\frac{h_{n}^{t}p_{n}^{t}}{N_{0}})} \\
\text{s.t.}  \quad & p_{n}^{\min} \leq p_{n}^{t} \leq p_{n}^{\max},\quad \forall n\in[N]. 
\end{align*}
The optimal solutions to the Problem~\textbf{P2.1.1} and~\textbf{P2.1.2} can be obtained through the following theorems:
%First, Since Problem~\eqref{prob_fnt} is convex, it can be solved by Theorem \ref{theo_solu_fnt}.
\begin{theorem}[Solution to \textbf{P2.1.1}]\label{theo_solu_fnt}
The optimal solution $(f_n^t)^*, \forall n \in[N]$ of \textbf{P2.1.1} can be obtained by
\begin{equation}\label{solu_fnt_opt}
(f_n^t)^* = \min\{ \max\{ \sqrt[3]{(f_n^t)^{\prime}} , f_{n}^{\min}\} , f_{n}^{\max} \},
\end{equation}
where $(f_n^t)^{\prime} = \frac{Vq_n^t}{Q_n^t (1-(1-q_{n}^{t})^{K})\alpha_{n}}$.
\end{theorem}
\begin{IEEEproof}
The main idea is to use the second-order derivation to prove convexity and use the first-order derivation to obtain the root. The proof details are provided in Appendix~\ref{append_theo_solu_fnt}.
\end{IEEEproof}

\begin{theorem}[Solution to \textbf{P2.1.2}]\label{theo_solu_punt}
The optimal solution $(p_{n}^t)^*, \forall n \in [N]$ of \textbf{P2.1.2} can be determined by:
\begin{equation}\label{eq_root_pnt_a5_main_1}
    (p_{n}^t)^* = \min\{ \max\{(p_{n}^t)^\prime, p_{n}^{\min}\} , p_{n}^{\max} \},
\end{equation}
where $(p_{n}^t)^\prime$ is the root of following equation:
\begin{equation*}\label{eq_root_pnt_a5_main}
    \ln{(1+\frac{h_{n}^tp_{n}^t}{N_0})} = \frac{h_{n}^tp_{n}^t + A_{1,n}N_0}{h_{n}^tp_{n}^t+N_0},
\end{equation*}
and $A_{1,n} = \frac{Vq_n^th_{n}^t}{Q_{n}^{t}(1-(1-q_n^t)^K)N_0}$.
\end{theorem}
%%%%%%%%%%%%%%%%%%%%
\begin{IEEEproof} 
% The core principle is utilizing the second-order derivative is positive. 
The main principle follows the same idea as \textbf{Theorem \ref{theo_solu_fnt}}. 
The details are summarized in Appendix~\ref{append_theo_solu_pnt}. 
\end{IEEEproof}

\subsubsection{Optimization w.r.t. $\bm{q}^t$}
Under fixed $\bm{f}^t$ and $\bm{p}^t$, we can reformulate the optimization problem $\textbf{P2}$ as
\begin{align*}
   % & \textbf{P3.2:} & & \notag \\
  \textbf{P2.2:} \quad \min_{\bm{q}^t}  \quad &f(\bm{q}^t)  =  V\sum_{n=1}^{N}(\mathcal{T}_{n}^{t}q_{n}^{t}+\lambda\frac{ w_n^2}{q_n^t}) \\
  & \quad \quad \quad ~~ - \sum_{n=1}^{N}\mathcal{E}_{n}^{t}(1-q_{n}^{t})^{K}\\
\text{s.t.} \quad  & \sum_{n=1}^{N}q_{n}^{t}=1, q_{n}^{t}\in (0,1], \forall n\in[N].  
\end{align*}

Note $f(\bm{q}^t)$ is the summation of one convex function and one concave function, which can be solved by successive upper-bound minimization (SUM) algorithm~\cite{razaviyayn2013unified}. For the simplicity, we define the convex function $f_{\text{cvx}}(\bm{q}^t)$ and concave function $f_{\text{cve}}(\bm{q}^t)$ as
\begin{align}
    f_{\text{cvx}}(\bm{q}^t) &\coloneqq \sum_{n=1}^{N}A_{2,n}q_{n}^{t}+\sum_{n=1}^{N}\frac{A_{3,n}}{q_n^t},\\
    f_{\text{cve}}(\bm{q}^t) &\coloneqq - \sum_{n=1}^{N}\mathcal{E}_n^t(1-q_{n}^{t})^{K},
\end{align}
where $A_{2,n} = V\mathcal{T}_{n}^{t}$ and $A_{3,n}=V\lambda w_n^2$.

Then, $\textbf{P2.2}$ can be rewritten as 
\begin{align*}
    \textbf{P2.2.1:} \quad  \min_{\bm{q}^t} \quad & f(\bm{q}^t) = f_{\text{cvx}}(\bm{q}^t)+f_{\text{cve}}(\bm{q}^t) \notag \\
    \text{s.t.} \quad & \sum_{n=1}^{N}q_{n}^{t}=1, q_{n}^{t}\in (0,1],\quad \forall n\in[N].   
\end{align*}

Following Example 3 in~\cite{scutari2016parallel}, we have the following convex upper approximation of $f(\bm{q}^t)$:
% To solve Problem \textbf{P3.2.1}, we define a surrogate function
\begin{equation*}
    g(\bm{q}^t, \bm{q}^{t, \tau}) \coloneqq f_{\text{cvx}}(\bm{q}^t) + (\bm{q}^t-\bm{q}^{t,\tau})^{T}\nabla f_{\text{cve}}(\bm{q}^{t,\tau}) + f_{\text{cve}}(\bm{q}^{t,\tau}).
\end{equation*}

Then, we use the SUM algorithm, which is guaranteed to converge as shown in Theorem 1 of~\cite{razaviyayn2013unified}. Specifically, the SUM algorithm first randomly initializes a feasible point $\bm{q}^{t,0}$. Then, it repeatedly solves $\bm{q}^{t,\tau+1}=\argmin_{\bm{q}^t}g(\bm{q}^t, \bm{q}^{t,\tau})$ by convex optimization tool, e.g., CVX~\cite{boyd2004convex},  until the convergence criterion $\|\bm{q}^{t,\tau+1}-\bm{q}^{t,\tau}\|_{2}\leq \epsilon$ is met, where $\epsilon$ is a given very small positive value. Finally, we have the solution $\bm{q}^*$, which satisfies that $\|\bm{q}^*-\bm{q}^{t,\tau}\|_{2}\leq \epsilon$. 
\begin{algorithm}[t]
\caption{Algorithm to Solve \textbf{P1}}\label{algorithm-2}
\textbf{Input:} $\{h_n^t, \forall n\}$, stopping condition $\epsilon_0, \epsilon_1$

\textbf{Ouput:} $\bm{f}^t=\{f_n^t, \forall n\}, \bm{p}^t= \{{q}_{n}^{t},\forall n\}, \bm{q}^{t}=\{ q_n^t, \forall n\}$

\begin{algorithmic}[1]
    \STATE Empirically initialize $\bm{f}_0^t, \bm{p}_0^t, \bm{q}_0^{t}$;\label{ln2:1}
    \STATE Set $\bm{z}_{0} \leftarrow (\bm{f}_0^t, \bm{p}_0^t, \bm{q}_0^{t})$ and $e \leftarrow 0$;\label{ln2:2}
    \WHILE{$\|\bm{z}_{e} - \bm{z}_{e-1}\|_2 > \epsilon_0$} 
    \STATE $\bm{f}_{e+1}^t \leftarrow$ solve equation~(\ref{solu_fnt_opt});\label{ln2:3}
    \STATE $\bm{p}_{e+1}^t \leftarrow$ solve equation~(\ref{eq_root_pnt_a5_main_1});\label{ln2:4}
    \STATE $\bm{q}_{e}^{t,0} \leftarrow \bm{q}_{e}^{t}$ and set $\tau \leftarrow 0$;\label{ln2:5}
    \WHILE{$\|\bm{q}_{e}^{t,\tau}-\bm{q}_{e}^{t,\tau-1}\|_{2} > \epsilon_{1}$}
    \STATE $\bm{q}_{e}^{t,\tau+1}\leftarrow\argmin_{\bm{q}_{e}^t}g(\bm{q}_{e}^t, \bm{q}_{e}^{t,\tau})$; \label{ln2:6}
    \STATE $\tau \leftarrow \tau+1$;\label{ln2:7}
    % \STATE $\bm{q}^{t,\tau+1} \leftarrow \bm{q}^{t,*}$
    % \STATE Solve $\bm{q}^t$ using SUM algorithm.
    \ENDWHILE
    \STATE $\bm{q}_{e+1}^{t} \leftarrow \bm{q}_{e}^{t,\tau+1}$;\label{ln2:8}
    \STATE $\bm{z}_{e+1} \leftarrow (\bm{f}_{e+1}^t, \bm{p}_{e+1}^t, \bm{q}_{e+1}^{t})$ and $e\leftarrow e+1$;\label{ln2:9}
    \ENDWHILE
    \STATE $\bm{f}^t, \bm{p}^t, \bm{q}^{t}\leftarrow\bm{f}_{e+1}^t, \bm{p}_{e+1}^t, \bm{q}_{e+1}^{t}$;
    \STATE Update $\bm{Q}^t$ by equations~(\ref{eq_virt_queue}) and (\ref{equ:ant});\label{ln2:10}
\end{algorithmic}
\end{algorithm}

We summarize the overall procedure in \textbf{Algorithm \ref{algorithm-2}}, which consists of two loops. The outer loop optimizes $\bm{f}^t, \bm{p}^t$, while the inner loop optimizes $\bm{q}^t$. In particular, we set the iteration index of the outer loop as $e$. At time slot $t$, the server collects the online observed channel statistic $h_n^t$ from all edge devices and uses them as algorithm input. Then we empirically make an initial guess on decision variables, e.g., $\{\bm{f}_0^t\}_n=(f_{n}^{\max}+f_{n}^{\min})/2, \{\bm{p}_0^t\}_{n}=(p_{n}^{\max}+p_{n}^{\min})/2, \{\bm{q}_0^{t}\}_{n}=1/N$ (line~\ref{ln2:1}). Next, we start the greedy algorithm to find the optimal decisions. Specifically, we first solve the equations (\ref{solu_fnt_opt}) and (\ref{eq_root_pnt_a5_main_1}) under fixed $\bm{q}_{e}^{t}$ to obtain $\bm{f}_{e+1}^t$ and $\bm{p}_{e+1}^t$ (lines~\ref{ln2:3}-\ref{ln2:4}). 

After obtaining $\bm{f}^t, \bm{p}^t$ through the outer loop, we start the inner loop optimization. We set the iteration index of the inner loop as $\tau$ and employ the SUM algorithm to solve the $\bm{q}_{e+1}^t$ under fixed $\bm{f}_{e+1}^t$ and $\bm{p}_{e+1}^t$ (lines~\ref{ln2:5}-\ref{ln2:8}). We iteratively solve $g(\bm{q}_{e}^t, \bm{q}_{e}^{t,\tau})$ (line~\ref{ln2:6}) until $\bm{q}_{e}^{t, \tau}$ meets the criteria $\epsilon_1$. The overall search process terminates if solutions meet the criteria $\epsilon_0$. Finally, we update the value of the virtual queue $\bm{Q}^t$ (line~\ref{ln2:10}).
%-----------------------------------------------------
\subsection{Performance Analysis}
%-----------------------------------------------------
In this subsection, we analyze the performance of LROA when the channel gains $h_{n}^t, \forall n$ are IID stochastic processes. Note that our results can be extended to the more general setting where $h_{n}^t,\forall n$ evolves according to some finite state irreducible and aperiodic Markov chains according to the Lyapunov optimization framework~\cite{neely2010stochastic}. 
%%%%%%%%%%%%%%%%%%%%%%%%%%
\begin{theorem}
If $h_{n}^t, \forall n$ are IID over time slot $t$, then the time-average expected objective value under our algorithm is within bound $C/V$ of the optimal value, i.e.,
\begin{multline}
\mathop{\textup{lim sup}}_{T\rightarrow\infty} \frac{1}{T}\sum_{t=0}^{T-1}\mathbb{E}\left\{\sum_{n=1}^{N} \left(q_{n}^{t}\mathcal{T}_{n}^{t}+\lambda \cdot  \frac{w_n^2}{ q_{n}^{t}}\right)\right\}\\
\leq \textbf{\textup{P2}}^*+C/V
\end{multline}
where C is the constant given in Lemma~\ref{lemma_drift_plus_penalty}.
\end{theorem}
%%%%%%%%%%%%%%%%%%%%%%%%%
\begin{IEEEproof}
As we mentioned before, our algorithm is always trying to greedily minimize the R.H.S of the upper bound~\eqref{in_drift_plus_penalty_upper_bound} of the \textit{drift-plus-penalty} term in each time slot $t$ over all possible feasible control decisions. Therefore, by plugging this policy into the R.H.S of the inequality~\eqref{in_drift_plus_penalty_upper_bound}, we have
\begin{align*}
\Delta_{V}(t)&\leq C+V\mathbb{E}[\sum_{n=1}^{N} \left(q_{n}^{t}\mathcal{T}_{n}^{t}+\lambda\frac{w_n^2}{ q_{n}^{t}}\right)|\bm{Q}^t]\\
& \leq C+V\textbf{\textup{P2}}^{*}.
\end{align*}
Summing over $t=0,\dots,T-1$, we have
\begin{multline*}
    V\mathbb{E}\sum_{t=0}^{T-1}[\sum_{n=1}^{N} \left(q_{n}^{t}\mathcal{T}_{n}^{t}+\lambda \cdot  \frac{w_n^2}{ q_{n}^{t}}\right)|\bm{Q}^t]\\
  \leq CT+VT\textbf{\textup{P2}}^{*}-\mathbb{E}[L(T)]+\mathbb{E}[L(0)].
\end{multline*}
Diving both sides by $T$, let $T\rightarrow\infty$ and using the facts that $\mathbb{E}[L(0)]$ are finite and $\mathbb{E}[L(T)]$ are nonnegative, we arrive at the following performance guarantee:
\begin{align}
    \mathop{\textup{lim sup}}_{T\rightarrow\infty}\ \frac{1}{T}\sum_{t=0}^{T-1}\mathbb{E}\left\{\sum_{n=1}^{N} \left(q_{n}^{t}\mathcal{T}_{n}^{t}+\lambda \cdot  \frac{w_n^2}{ q_{n}^{t}}\right)\right\}
\leq \textbf{\textup{P2}}^*+C/V,
\end{align}
where $\textbf{\textup{P2}}^*$ is the optimal objective value, $C$ is a constant, and $V$ is a control parameter.
\end{IEEEproof}

%-----------------------------------------------------
\section{Experiments}\label{sec:exp}
%-----------------------------------------------------
% \red{(The experimental results are unfinished yet.)}

% To demonstrate the effectiveness of our proposed algorithm, we perform a series of experiments under diverse resource capabilities and data distributions. 
% In this section, we first introduce the experiment setup and baselines. Subsequently, we demonstrate the experiment results and validate our theoretical conclusions in Section \ref{sec:analysis}.

\subsection{Experiment Setup}
% \red{(See https://arxiv.org/abs/1602.05629 or our CFEL paper for dataset description.)}
% Following~\cite{nguyen2020efficient,yang2020energy,vu2020cell}, the network environment is set as follows by default.
We emulate a large number of devices in a GPU server and use real-world measures to set the configurations. Specifically, we consider a FL system with 120 edge devices and one server. Referring to the setting in~\cite{nguyen2020efficient,yang2020energy,vu2020cell}, unless explicitly specified, the default configurations of the FL system are as follows. Each edge device has the maximum transmission power $p_{n}^{\text{max}}= \SI{0.1}{W}$ and minimum transmission power $p_{n}^{\text{min}}= \SI{0.001}{W}$. The white noise power spectral density is $N_{0}=\SI{0.01}{W}$. The maximum available CPU frequency $f_{n}^{\text{max}}$ is $\SI{2.0}{GHz}$ and the minimum CPU frequency $f_{n}^{\text{min}}$ is $\SI{1.0}{GHz}$. The efficient capacity coefficient of CPU is $\alpha_{n} = 2\times 10^{-28}$. 
% In practical edge devices, the download link (from the server to edge devices) usually has a larger bandwidth than the uplink (from edge devices to the server). 
% The download time is not bottleneck compared with the upload time. 
For simplicity, we ignore download cost and only consider upload time in the experiment. The total uplink communication bandwidth $B = \SI{1}{MHz}$. To simulate the practical communication environment, we generate the random channel gain following the exponential distribution with a mean value of $\bar{h}_n^t=0.1$. Note we fix the random seed of random channel gain across different runnings. Moreover, we filter out the outlier greater than 0.5 or smaller than 0.01 to ensure the generated random channel gain is within a reasonable range. 
% In our first step of evaluation, we only consider the heterogeneity in communication channels.

% 
% \hl{We can find the default system setting is a homogeneous system.} \red{(What do you mean by homogeneous system? The key idea behind our adaptive scheme is heterogeneity. If homogeneous, why using adaptive sampling?)}

We use two image classification datasets: FEMNIST~\cite{caldas2019leaf} and CIFAR-10~\cite{krizhevsky2009learning} for experiment. 
% The FEMNIST dataset is a federated splitting version of the EMNIST dataset that consists of handwritten letters and digits (62 classes) from 3500 different writers. 
Due to the inherent diversity in writing styles among the writers, the FEMNIST is a naturally non-IID distributed dataset. Following~\cite{caldas2018leaf}, we first filter out the writers who contribute less than 50 samples. Subsequently, we randomly pick 120 writers from the remaining pool to simulate the devices in the FL system. 
% We reserve $10\%$ of their respective samples from each edge device for testing purposes. Ultimately, a common testing dataset is constructed by pooling the testing data from all edge devices. 
For CIFAR-10 dataset, the 50,000 training images are divied into 120 devices following Dirichlet distribution \cite{hsu2019measuring} with concentration parameter 0.5. We train a ResNet-18~\cite{he2016deep} ($d=11,172,342$ total parameters) for CIFAR-10, and a CNN model with $d=6,603,710$ total parameters for FEMNIST. 
% Following~\cite{hsieh2020non}, we replace batch normalization with group normalization in ResNet-18 for better performance. 

% , which consists of two convolutional layers followed by two fully connected layers. The convolutional layers have 32 and 64 channels, respectively. The convolutional kernel size is 5, and the fully connected layer has 2,048 hidden neurons. Finally, a final softmax output layer is appended after the fully connected layer. The CNN model has . 

% We evenly divide the 50,000 training images into 120 devices in an IID fashion. 
% The 50,000 training images are divided into 120 partitions in an IID fashion.in an IID fashion. 
% \hl{By doing so, we first exempt the influence of data heterogeneity and can evaluate the performance of LROA when facing system heterogeneity.} 
% The 10,000 testing images are used as the common testing data. We train a ResNet-18~\cite{he2016deep} ($d=11,172,342$ total parameters) for CIFAR-10. Following~\cite{hsieh2020non}, we replace batch normalization with group normalization in ResNet-18 for better performance. 

We calculate the model update size as $M=32\times d$ since the model parameter is represented by a 32-bit floating point number. Server samples $K=2$ times to create the selected devices set at the beginning of each communication round. The number of CPU cycles required to process a data sample is $c_{n}=2.0\times 10^{9} \text{ cycles/sample}$ for FEMNIST and $c_{n}=3.0\times 10^{9} \text{ cycles/sample}$ for CIFAR-10. The available energy supply is $\bar{\mathcal{E}}_{n}=\SI{5}{J}$ for FEMNIST, and $\bar{\mathcal{E}}_{n}=\SI{15}{J}$ for CIFAR-10. The total training rounds are 2000 and 1000 for CIFAR-10 and FEMNIST, respectively. Each device performs $E=2$ local epochs at each round. We adopt the SGD optimizer with the momentum of 0.9 to update the local model, where the initial learning rate is 0.05 for CIFAR-10 and 0.1 for FEMNIST. The learning rate is decayed by half at $50\%$ and $75\%$ of the total training rounds. We run each experiment 30 times using different random seeds and present the average testing accuracy for comparison. 

% has maximum energy consumption budget per round $\bar{\mathcal{E}}_{n}=15 \text{ J}$, and $\bar{\mathcal{E}}_{n}=5 \text{ J}$ for FEMNIST.}

To show the effectiveness of LROA, we compared it with FedAvg under uniform sampling and vanilla resource allocation. Specifically, we consider the following two baselines:
\begin{itemize}

    \item \emph{Uniform Sampling with Dynamic Resource Allocation (Uni-D)}: At the beginning of each communication round, the server samples $K$ times using $q_n^t=1/N$ to build the selected edge device set $\mathcal{K}^t$. Besides, we still use LROA to find the optimal $p_n^t$ and $f_n^t$ for edge devices.
    
    % To satisfy the energy constraint, we still use LROA to find the optimal $p_n^t$ and $f_n^t$ while fixing $q_n^t=1/N$. By comparing this approach with LROA, we prove the contribution of adaptive client sampling.
    
    % For simplicity, we keep $\lambda$ and $V$ the same as our adaptive sampling case. 
    
    \item \emph{Uniform Sampling with Static Resource Allocation (Uni-S)}: At the beginning of each communication round, the server samples $K$ times using $q_n^t=1/N$ to construct the selected set of edge device set $\mathcal{K}^t$. For edge device's communication power and computation frequency, we assume the communication power operates at the mid-level and computation consumes the remaining energy. Specifically, we set the transmission power $p_{n}^{t}=(p_n^{\text{min}}+p_n^{\text{max}})/2$ and the computation frequency of each device satisfying $[E\alpha_n c_n D_n (f_{n}^{t})^2/2+p_{n}^{t}MK/{(B\log_2(1+h_n^t p_{n}^{t}/N_0))}](1-(1-{1}/{N})^{K})=\bar{\mathcal{E}}_n$. If the solution of $f_n^t$ falls outside the feasible region, we project it to the nearest boundary value. 

    \item \emph{Diverse Client Selection via Submodular Maximization (DivFL)}: DivFL~\cite{balakrishnan2022diverse} is prior state-of-art method in optimizing the client selection for FL. Specifically, it optimally selects a small diverse subset of clients who carry the representative gradient information. By updating the subset to approximate full updates through the aggregation of all client information, it achieves a better convergence rate compared to random selection. As DivFL focuses on client selection solely, we adapt it to our problem setting by computing the transmission power and the computation frequency using the same strategy with \emph{Uni-S}.

    % To satisfy the energy constraint, this baseline considers a static strategy and chooses the resource allocation decisions such that the resulting energy consumption equals the energy rate limit $\bar{\mathcal{E}}_n$. Specifically, we set the transmission power $p_{n}^{t}=\frac{1}{2}(p_n^{\text{min}}+p_n^{\text{max}})$ and the computation frequency of each device satisfying $[E \frac{\alpha_n}{2} c_n D_n (f_{n}^{t})^2+\frac{p_{n}^{t}MK}{(B\log_2(1+h_n^t p_{n}^{t}/N_0))}](1-(1-\frac{1}{N})^{K})=\bar{\mathcal{E}}_n$. If the solution of $f_n^t$ falls outside the feasible region, we project it to the nearest boundary value. 
    
    % Server aggregates the local model updates with weighted average $\bm{\theta}^{t+1} \xleftarrow[]{} \sum_{n\in\mathcal{K}^t}\frac{w_n}{\sum_{n\in\mathcal{K}^t}w_n}\bm{\theta}_n^{t,E}$.
    
    % \item \emph{Weighted Sampling}: In every global round, server samples $K$ times using probability $q_{n}^{t} = w_{n}$ to determine $\mathcal{K}^{t}$. Server aggregates the local model updates with weighted average. To satisfy the energy constrain, we set the communication power of each device as the solution of equation $\frac{p_{n}^{t,\text{wei}}(1-(1-w_n)^{K})}{\log_{2}(1+h_{n}^{t} p_{n}^{t,\text{wei}}/N_{0})} = \frac{p_{n}^{\text{max}} (1-(1-1/N)^{K})}{\log_{2}(1+h_{n}^{t} p_{n}^{\text{max}}/N_{0})}$, the computation power of each device is $f_{n}^{t,\text{wei}} = f_{n}^{\text{max}}\sqrt{\frac{1-(1-\frac{1}{N})^{K}}{1-(1-w_n)^{K}}}$. 
    
\end{itemize}

The baselines are chosen to mimic the solutions proposed in the recent references when adapted to our unique problem setting. Specifically, (\emph{Uni-D}) is adopted from~\cite{yang2020energy,tran2019federated}, which focuses on optimizing resources under the uniform sampling. It serves to demonstrate the effectiveness of the obtained probability vector $q_n^t$. The comparison between \emph{Uni-D} and LROA highlights the effectiveness of adaptive client sampling. (\emph{Uni-S}) represents a more standard or 'vanilla' solution. It assumes static resource allocation and uniform client sampling. The comparison between \emph{Uni-D} and \emph{Uni-S} proves the benefits of resource optimization.

% \hl{For uniform sampling, the server aggregates the local model updates with weighted average $\bm{\theta}^{t+1} \xleftarrow[]{} \sum_{n\in\mathcal{K}^t}\frac{w_n}{\sum_{n\in\mathcal{K}^t}w_n}\bm{\theta}_n^{t,E}$.} \red{(why do we need this?)}

% All the algorithms are developed using PyTorch and executed on an Ubuntu server with 4 NVIDIA RTX A6000 GPUs.
%-------------------------------------
\subsection{Experiment Results}
%-------------------------------------
%%%%%%%%%%%%%
\subsubsection{Comparing to Baselines}

\begin{figure}[t]\centering
    % \subfloat[]{{\includegraphics[width=0.22\textwidth]{ {./figure/cifar_train_loss_baselines.pdf}} }}
    \subfloat[]{{\includegraphics[width=0.22\textwidth]{ {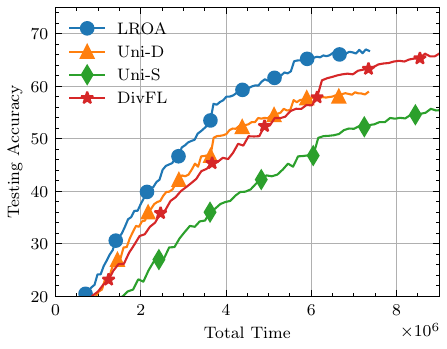}} }} 
    % \\
    % \subfloat[]{{\includegraphics[width=0.22\textwidth]{ {./figure/cifar_train_loss_baselines_round.pdf}} }}
    \subfloat[]{{\includegraphics[width=0.22\textwidth]{ {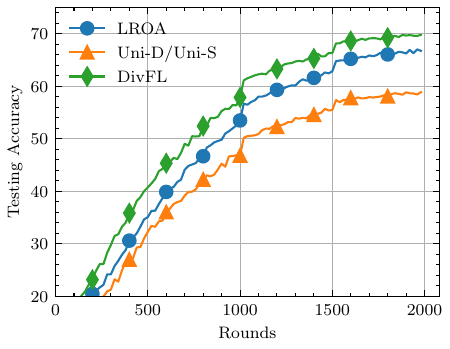}} }}
    \caption{Convergence rate and runtime comparisons of LROA and baselines on CIFAR-10. (a) Testing accuracy with runtime; (b) Testing accuracy with communication rounds. 
    % \red{(Training loss should be the left, and testing accuracy should be the right.)}
    }
    \label{fig:baseline_cifar}%
\end{figure}
\begin{figure}[t]\centering
    % \subfloat[]{{\includegraphics[width=0.22\textwidth]{ {./figure/femnist_train_loss_baselines.pdf}} }}
    \subfloat[]{{\includegraphics[width=0.22\textwidth]{ {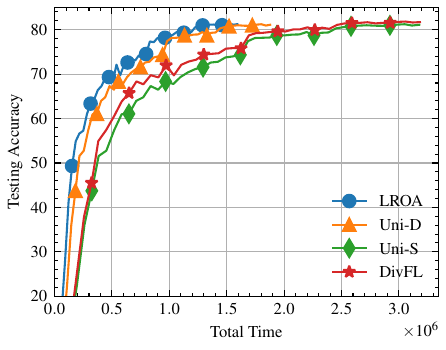}} }}
    % \subfloat[]{{\includegraphics[width=0.22\textwidth]{ {./figure/femnist_train_loss_baselines_round.pdf}} }}
    \subfloat[]{{\includegraphics[width=0.22\textwidth]{ {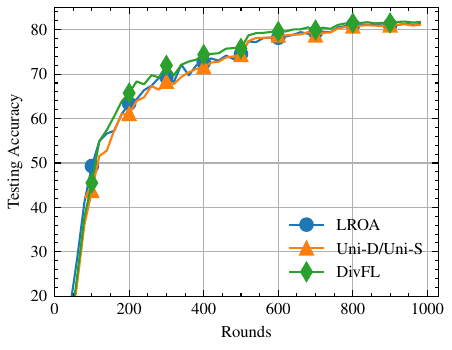}} }}
    \caption{Convergence rate and runtime comparisons of LROA and baselines on FEMNIST. (a) Testing accuracy with runtime; (b) Testing accuracy with communication rounds.}
    \label{fig:baseline_femnist}%
\end{figure}

We begin by comparing LROA with the baselines regarding the convergence speed and total running time. $\lambda$ and $V$ are two hyper-parameters that need careful choosing. To ensure a reasonable starting point for $\lambda$ and $V$, we estimate $\lambda_{0}$ as $\mathcal{T}_{0}/{F_{0}}$, where $\mathcal{T}_{0}$ represents the estimated per-round time consumption when $p_{n}^{t}=(p_n^{\text{min}}+p_n^{\text{max}})/2$ and $f_{n}^{t}=(f_n^{\text{min}}+f_n^{\text{max}})/2$, and $F_{0}$ is estimated loss by fixing $q_{n}^{t}=w_{n}^{t}$. During experiments, we set $\lambda=\mu \lambda_{0}$, where $\mu$ serves as a scaling factor of $\lambda$. Once we have a specific value for $\lambda$, we estimate $V_{0}$ as ${Q_0a_{0}}/{(\mathcal{T}_{0} + \lambda F_{0})}={a_{0}^{2}}/{(\mathcal{T}_{0} + \lambda F_{0})}$ where $a_{0}$ denotes the estimated of energy reminds by computing equation (\ref{equ:ant}) (We also estimate virtual query as $Q_0=a_0$). Finally, we set $V=\nu V_{0}$, where $\nu$ acting as the scaling factor for $V_0$.

% \red{(Figure caption is wrong. Should be LROA, not LORA.)}

%-------------------------------------

% FEMNIST
% our: 1623128.82
% Uni-D: 1915775.91
% Uni-S: 3238961.29

% CIFAR
% our: 2984604.84
% Uni-D: 3770272.57
% Uni-S: 5979172.95

% \hl{We fix $\mu = 1.0$ and $\nu = 10^{5}$ for LROA and \emph{Uni-D} for CIFAR-10 and FEMNIST.} 
% \red{($\lambda$ should be the same across all algorithms. $V$ can be tuned for adaptive resource allocation algorithms. Only need to spell out what $V$ and $\lambda$ values are selected when drawing each figure.)}

The testing accuracy w.r.t. running time and communication round for CIFAR-10 and FEMNIST are shown in Fig.~\ref{fig:baseline_cifar} and Fig.~\ref{fig:baseline_femnist}, respectively. Here we fix $\mu = 1.0$ and $\nu = 10^{5}$ for CIFAR-10 and FEMNIST. Note we set the same $\mu, \nu$ for \emph{Uni-D} for easy comparison. In order to provide a clearer representation of the ending points of each curve, we have included a solid circle at the termination of each curve. From the figures, we can observe LROA can achieve better testing accuracy in terms of total training latency. Specifically, LROA saves $20.8\%$ and $50.1\%$ total training time compared with \emph{Uni-D} and \emph{Uni-S} for CIFAR-10, and $15.3\%$ and $49.9\%$ for FEMNIST.
% \red{(Fill out the correct number. Given a target accuracy, what is the running time comparison?)}

% \red{(The following part remains to be checked...)}

% \st{While our algorithm only has minor degradation compared with uniform sampling in terms of communication rounds. We also notice in terms of convergence speed with respect to rounds, LROA is slightly better than uniform sampling at the beginning stage of training. This is because FEMNIST is an unbalance FL dataset. Solving $q_n^t$ with LROA forces the server to pick the device with more data samples. As a result, the convergence speed is faster than uniform sampling in terms of rounds.}

Note \emph{Uni-D} adopts the same strategy to assign the communication power and CPU frequency as LROA, except for client sampling. Therefore, the enhancement observed from \emph{Uni-D} to LROA demonstrates the effectiveness of adaptive client sampling. On the other hand, \emph{Uni-S} employs fixed communication power and CPU frequency. Consequently, the improvement from \emph{Uni-S} to \emph{Uni-D} confirms the benefit of control in communication power and CPU frequency. 
%%%%%%%%%%%%%%%%%%%%%%
\subsubsection{Comparing Different Configurations of $\lambda$ and $V$}
\begin{figure}[t]
\centering
    % \subfloat[CIFAR-10]{{\includegraphics[width=0.22\textwidth]{ {./figure/cifar_train_loss_lambda.pdf}} }}
    \subfloat[CIFAR-10]{{\includegraphics[width=0.22\textwidth]{ {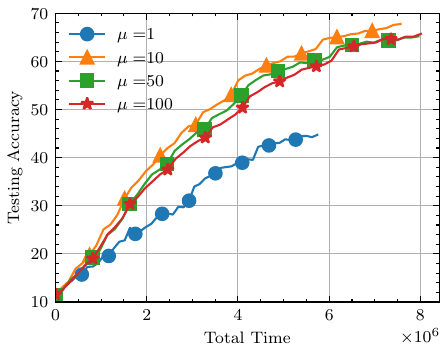}} }} 
    % \subfloat[FEMNIST]{{\includegraphics[width=0.22\textwidth]{ {./figure/femnist_train_loss_lambda.pdf}} }} 
    \subfloat[FEMNIST]{{\includegraphics[width=0.22\textwidth]{ {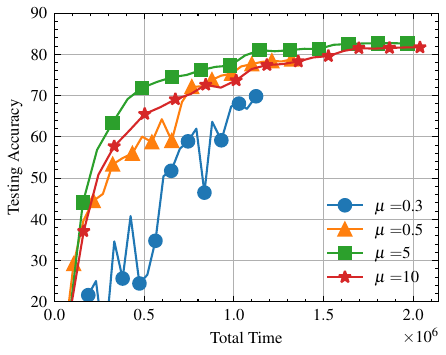}} }}
    \caption{Testing accuracy vs. total time of LROA under different $\lambda$ on CIFAR-10 (a) and FEMNIST (b). 
    % (a,c) Training loss with runtime. (b,d) Testing accuracy with runtime.
    }
    \label{fig:lambda}%
\end{figure}

%%%%%%%%%%%%%%%%%%%%%%%
We proceed by examining the impact of the parameter $\lambda$. The value of $\mu$ is varied across $\{1.0, 10.0, 50.0, 100.0\}$ for CIFAR-10 dataset, and $\{0.3, 0.5, 5.0, 10.0\}$ for FEMNIST dataset. The value of $\nu$ fixed at $10^{5}$ for both datasets. In Fig.~\ref{fig:lambda}, we compare the performance of LROA with baseline methods. The results illustrate that as $\lambda$ increases, the total time cost required to complete $T$ rounds also increases. Furthermore, higher values of $\lambda$ lead to improved testing accuracy. It is important to note that when $\lambda$ approaches zero, the contribution of the loss function in the objective of Problem \textbf{P2} tends to diminish, which means resource allocation solely. Observing the figure, we can deduce that a smaller $\mu$ yields a highly volatile curve, particularly noticeable when $\mu=0.3$, which exhibits significant instability. This observation emphasizes the necessity of jointly considering model training and resource control.

% Next, we investigate the influence of $\lambda$. We vary the $\mu$ from $\{0.3, 0.5, 5.0, 10.0\}$ while fixing $\nu=10^{5}$ and compare the performance of LROA with baselines in Fig.\ref{fig:lambda}. As shown in the figure, larger $\lambda$ needs a longer training time to finish total $T$ rounds, the testing accuracy is better, and the training loss is smaller. Note when $\lambda$ goes to zero, which means the contribution of the loss function in the objective of P(2) goes to zero. From the figure, we can find a smaller $\mu$ leads to a highly fluctuated curve, especially when $\mu=0.3$, the curve is highly unstable. This confirms the necessity of joint considering model training and resource control.  

%($\mu: 0.3 \rightarrow 0.5 \rightarrow 5.0$), then higher $\lambda$ yields a worse curve ($\mu: 5.0 \rightarrow 10.0$). \hl{The result shows an interesting trade-off between training performance and running time. }

\begin{figure}[t]\centering
    \subfloat[CIFAR-10]{{\includegraphics[width=0.22\textwidth]{ {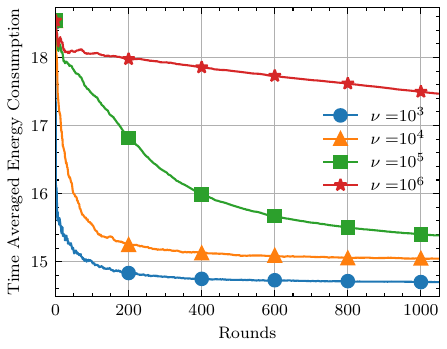}} }} 
    \subfloat[CIFAR-10]{{\includegraphics[width=0.22\textwidth]{ {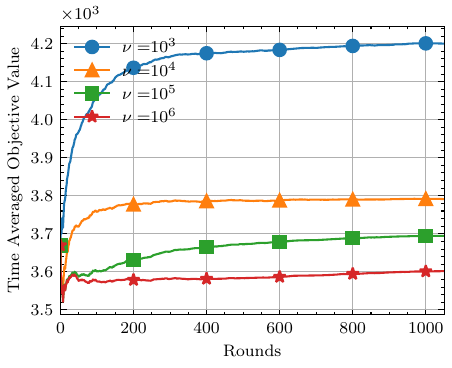}} }}\\
    \subfloat[FEMNIST]{{\includegraphics[width=0.22\textwidth]{ {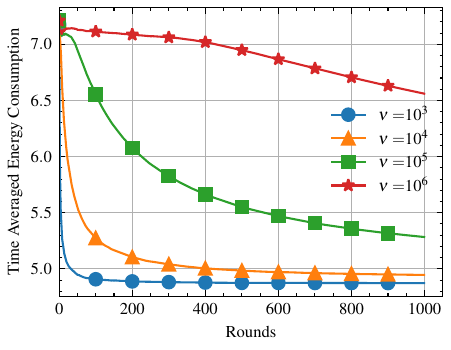}} }} 
    \subfloat[FEMNIST]{{\includegraphics[width=0.22\textwidth]{ {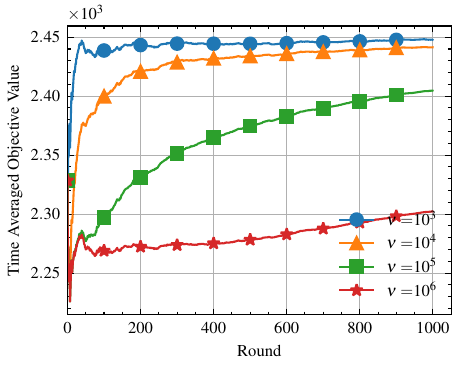}} }}
    \caption{Convergence of energy and objective value for CIFAR-10 (a,b) and FEMNIST (c,d). (a,c) Expected time-averaged energy consumption. (b,d) Expected time averaged objective value. $\mu=1.0$ and $\nu\in \{10^3,10^4,10^5,10^6\}$. }
    \label{fig:V}%
\end{figure}

Next, we study the impact of $V$. In Fig.~\ref{fig:V}, we plot the time-averaged energy consumption $\sum_{t=0}^{T-1}(1-(1-q_n^t)^K)(\mathcal{E}_{n}^{t,\text{com}}+\mathcal{E}_{n}^{t,\text{cmp}})/T$ and the time-averaged objective value $(q_n^t\mathcal{T}_n^t+\lambda{w_n^2}/{q_n^t})/T$. Note the curves are averaged across all devices here. We vary $\nu \in \{10^3, 10^4, 10^5, 10^6\}$. As shown in Fig.~\ref{fig:V}(a), larger $\nu$ results in a slower convergence rate target to energy budget, which means bad satisfaction with energy constraints. Note the energy budget is $\bar{\mathcal{E}}_n = \SI{15}{J}$ for CIFAR-10, $\bar{\mathcal{E}}_n = \SI{5}{J}$ for FEMNIST. From Fig.~\ref{fig:V}(b), we can find that large $V$ results in a small time-averaged objective value, which means good optimization results in terms of running time. Clearly, $V$ controls the trade-off between objective minimization and constraint satisfaction as stated in Section \ref{sec:lya-app}.

\begin{figure}[t]\centering
    % \subfloat[CIFAR-10]{{\includegraphics[width=0.22\textwidth]{ {./figure/cifar_train_loss_K24816.pdf}} }}
    \subfloat[CIFAR-10]{{\includegraphics[width=0.22\textwidth]{ {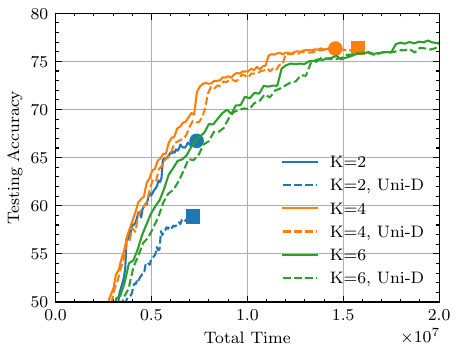}} }}
    % \subfloat[FEMNIST]{{\includegraphics[width=0.22\textwidth]{ {./figure/femnist_train_loss_K24816.pdf}} }}
    \subfloat[FEMNIST]{{\includegraphics[width=0.22\textwidth]{ {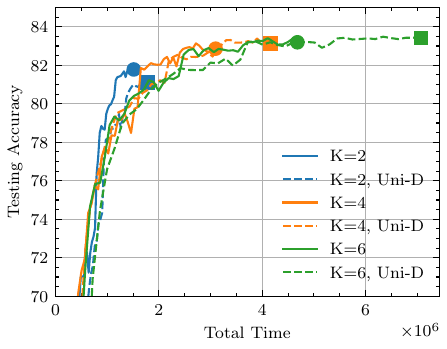}} }} 
    \caption{Testing accuracy vs. total time of LROA under different sampling numbers $K$ on CIFAR-10 (a) and FEMNIST (b). The marker shows the ending point of each curve. (Circle: LROA, Square: \emph{Uni-D})}
    \label{fig:K}%
\end{figure}
\begin{figure}[t]\centering
    % \subfloat[CIFAR-10]{{\includegraphics[width=0.22\textwidth]{ {./figure/cifar_train_loss_K24816_round.pdf}} }}
    \subfloat[CIFAR-10]{{\includegraphics[width=0.22\textwidth]{ {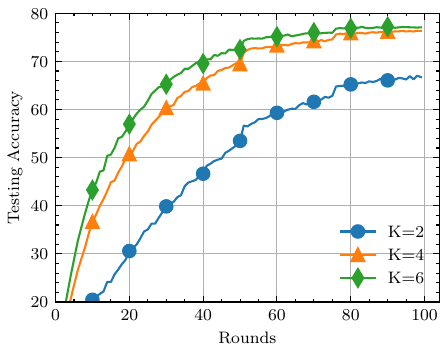}} }}
    % \subfloat[FEMNIST]{{\includegraphics[width=0.22\textwidth]{ {./figure/femnist_train_loss_K24816_round.pdf}} }}
    \subfloat[FEMNIST]{{\includegraphics[width=0.22\textwidth]{ {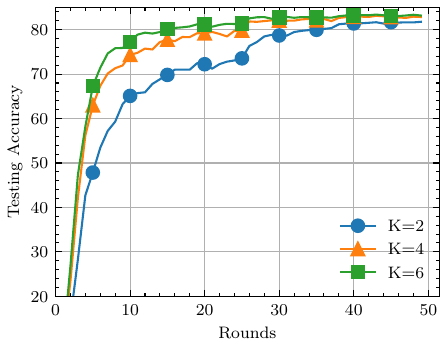}} }} 
    \caption{Testing accuracy vs. communication round of LROA under different sampling numbers $K$ on CIFAR-10 (a) and FEMNIST (b).}
    \label{fig:K_round}%
\end{figure}
%%%%%%%%%%%%%%%%%%%%%%%%%
\subsubsection{Impact of $K$}
%%%%%%%%%%%%%%%%%%%%%%%%%
We evaluate the performance of LROA across different server sampling frequencies $K$. The range of $K$ is varied to $\{2, 4, 6 \}$. As demonstrated in the previous experiment, the hyperparameters $\lambda$ and $V$ play a crucial role in surpassing the performance of baselines. For each $K$, we conduct a grid search to identify the optimal values within range $\mu \in \{0.1, 1.0, 10.0\}$ and $\nu \in \{10^4,10^5,10^6\}$. It should be noted that the corresponding values of $\lambda_{0}$ and $V_{0}$ are recalculated accordingly. We select the hyperparameter combination that hits the best time-accuracy trade-off to represent the true performance of LROA.

Specifically, since both the total running time and the final testing accuracy are critical in evaluating our algorithm, we aim to strike a balance between them. We first filter out the curves that do not meet the testing accuracy criteria. Then, we sort the remaining curves based on their total running time and select the one with the minimum value as the optimal choice. In our experiments, we define the filter-out criteria for $K\in\{2,4,6\}$ as $\{64, 76, 77\}$ for CIAFR-10 and $\{80, 81, 82\}$ for FEMNIST.

The previous experiment results have already demonstrated that the \emph{Uni-D} model consistently outperforms the \emph{Uni-S} and \emph{DivFL}. Therefore, for the sake of simplicity, we only display the results for the \emph{Uni-D}. To ensure a fair comparison, we also search for the optimal values of $\mu$ and $\nu$ for the \emph{Uni-D} model within the same range, following the same principles. The results are depicted in Fig.~\ref{fig:K} and~\ref{fig:K_round}.

From the figures, we can observe that our algorithm consistently outperforms the baseline algorithm across various sampling numbers $K$. This further validates the superiority of our proposed algorithm. Additionally, we notice that as $K$ increases, more communication time is required. The reason is that with more devices joining the learning, less communication bandwidth is available to a single device. Another possible reason is that, under the default setting, all devices are assumed to have the same communication and communication resources, except for different communication channels. Our problem becomes a channel selection problem. As more devices are selected, there is a higher likelihood of selecting a device with a poor channel. Furthermore, we observe that the final testing accuracy improves with increasing $K$, which aligns with our theoretical analysis presented in Theorem~\ref{th:convergence}.

\bibliographystyle{IEEEtran}
\bibliography{IEEEabrv,IEEE}

\appendices
\section{Proof of Equation(~\ref{equ:aggre})}\label{append_agg}
We take the expectation for the aggregated model $\bm{\theta}^{t+1}$ over the randomness of client sampling, we have
\begin{align}
    &\mathbb{E}_{\mathcal{K}^{t}} [ \bm{\theta}^{t+1}]  = \mathbb{E}_{\mathcal{K}^{t}} [\bm{\theta}^{t} + \sum_{n\in\mathcal{K}^t}\frac{w_n}{Kq_n^t}(\bm{\theta}_n^{t,E}-\bm{\theta}^{t}) ]\notag\\
    % & = \bm{\theta}^{t} + \mathbb{E}_{\mathcal{K}^{t}} \Big[ \sum_{n\in\mathcal{K}^t}\frac{w_n}{Kq_n^t}(\bm{\theta}_n^{t,E}-\bm{\theta}^{t}) \Big] \\
    & \labelrel={equ:expect} \bm{\theta}^{t} + K\sum_{n=1}^{N}q_n^t\frac{w_n}{Kq_n^t}(\bm{\theta}_n^{t,E}-\bm{\theta}^{t}) 
    % & = \bm{\theta}^{t} + K\sum_{n=1}^{N}\frac{w_n}{K}\bm{\theta}_n^{t,E} - K\sum_{n=1}^{N}\frac{w_n}{K}\bm{\theta}^{t}\\
    % & = \bm{\theta}^{t} + \sum_{n=1}^{N}w_n\bm{\theta}_n^{t,E} - \sum_{n=1}^{N}w_n\bm{\theta}^{t} \\
    = \sum_{n=1}^{N}w_n\bm{\theta}_n^{t,E} = \bm{\bar{\theta}}^{t+1}.\notag
\end{align}
Here $\bm{\bar{\theta}}^{t+1}$ denotes the weighted aggregation model under full client participation. \eqref{equ:expect} is derived by examining the contribution of individual clients. 
% For each sampling event, client $n$ has possibility $q_n^t$ to be selected, the underlying contribution is $q_n^t \cdot \frac{w_n}{Kq_n^t}(\bm{\theta}_n^{t,E}-\bm{\theta}^{t})$. 
As every client can be selected by the same probability $q_n^t$ and sampling $K$ times, 
% the total contribution from all clients is $\sum_{n=1}^{N}q_n^t\frac{w_n}{Kq_n^t}(\bm{\theta}_n^{t,E}-\bm{\theta}^{t})$. The total client sampling times are $K$. Thus, 
the expectation is $K\sum_{n=1}^{N}q_n^t\frac{w_n}{Kq_n^t}(\bm{\theta}_n^{t,E}-\bm{\theta}^{t})$. 

%%%%%%%%%%%%%%%%%%%%%%%%%%%%%%%%%%%%%%%%%%%%%%%%%%%%%%%%%%%%%%%%%%%%%%%%%%%%%%%%%%%%%%%%%%%%%%%%%%%%%%%%%%%%%%%%%%%%%%%%%%%%%%%%%%%%%%%%%%%%%%%%%%%%%%%%%%%%%%%%%%%%%%%%%%%%%%%%%%%%%%%%%%%%%%%%%
%--------------------------------
\section{Proof of Theorem 1}\label{append_conv}
%------------------------------------------------------
\subsection{Useful Lemmas}

\setcounter{lemma}{1}
% \begin{lemma}[Jensen's inequality]\label{lemma_jensen_in}
% For arbitrary set of $n$ vectors $\{\bm{a}_i\}_i^n$, $\bm{a}_i\in\mathbb{R}^d$ and positive weights $\{w_i\}_{i\in [n]}$, $\sum_{i}^nw_i=1$,
% \begin{equation*}
%     \norm{\sum_{i=1}^n w_i \bm{a}_i}^2 \leq \sum_{i=1}^n w_i\norm{\bm{a}_i}
% \end{equation*}
% \end{lemma}

% \begin{lemma}[Cauchy-Schwarz inequality]\label{lemma_cau_sch_in}
% For arbitrary set of $n$ vectors $\{\bm{a}_i\}_{i=1}^n$, $\bm{a}_i \in \mathbb{R}^d$,
% \begin{equation*}
%     \norm{\sum_{i=1}^n\bm{a}_i}^2\leq n\sum_{i=1}^n\norm{\bm{a}_i}^2
% \end{equation*}
% \end{lemma}

\begin{lemma}\label{lemma_in_square_sum}
For given two vectors $\bm{a},\bm{b}\in \mathbb{R}^d$,
\begin{equation*}
    \|\bm{a}+\bm{b}\|^2\leq(1+\alpha)\|\bm{a}\|^2+(1+\alpha^{-1})\|\bm{b}\|^2, \forall \alpha \geq 0.
\end{equation*}
\end{lemma}

% \begin{lemma}\label{lemma_in_inn_prod}
% For given two vectors $\bm{a},\bm{b}\in \mathbb{R}^d$,
% \begin{equation*}
%     2\langle \bm{a},\bm{b}\rangle \leq \gamma\norm{\bm{a}}^2 +\gamma^{-1}\norm{\bm{b}}^2, \forall \gamma \geq 0.
% \end{equation*}
% \end{lemma}

%%%%%%%%%%%%%%%%%%%%%%%%%%%%%%%%%%%%%%%%%%%%%%%%%%
\begin{lemma}[Unbiased sampling and Bounded expectation]\label{lemma_exp_mom_sampling}
Suppose the devices set $\mathcal{K}^t$ are sampled with probability $\bm{q}^t = [q_1^t,\dots, q_n^t]$ and their local models are $\bm{\theta}_n^t$, we have 
\begin{align*}
    \mathbb{E}_{\mathcal{K}^t}\sum_{n\in{\mathcal{K}^t}} \frac{w_n}{Kq_n}\bm{\theta}_n^t& =\sum_{n=1}^{N}w_n\bm{\theta}_n^t\\
    \mathbb{E}_{\mathcal{K}^t}\|\sum_{n\in{\mathcal{K}^t}}\frac{w_n}{Kq_n}\bm{\theta}_n^t\|^2 & \leq  \sum_{n=1}^{N}\frac{w_n^2}{Kq_n}\|\bm{\theta}_n^t\|^2
\end{align*}
\end{lemma}
%%%%%%%%%%%%%%%%%%%%%%%%%%%%%%%%%%%%%%%%%%%%%%%%
\begin{IEEEproof} 
Let $\mathcal{K}^t=\{n_1,n_2,\dots,n_K\}\subset[N]$. We have

\begin{align*}
    &\mathbb{E}_{{\mathcal{K}^t}}\sum_{n\in{\mathcal{K}^t}}\frac{w_n}{Kq_n}\bm{\theta}_n^t=\mathbb{E}_{{\mathcal{K}^t}}\sum_{n=1}^{K}\frac{w_{n_i}}{Kq_{n_i}}\bm{\theta}_{n_i}^t=K\mathbb{E}_{{\mathcal{K}^t}}[\frac{w_{n_1}}{Kq_{n_1}}\bm{\theta}_{n_1}^t]\\
    &= K\sum_{n=1}^{N}\frac{w_{n}}{Kq_{n}}q_{n}\bm{\theta}_{n}^t = \sum_{n=1}^Nw_{n}\bm{\theta}_n^t.
\end{align*}
For the expectation of square variables, we get
\begin{align*}
    &\mathbb{E}_{\mathcal{K}^t}\|\sum_{n\in{\mathcal{K}^t}}\frac{w_n}{Kq_n}\bm{\theta}_n^t\|^2 = \frac{1}{K^2}\mathbb{E}_{\mathcal{K}^t}\|\sum_{i=1}^{K}\frac{w_{n_i}}{q_{n_i}}\bm{\theta}_{n_i}^t\|^2\\
    & \labelrel\leq{in_exp_cauchy} \frac{1}{K}\mathbb{E}_{\mathcal{K}^t}\|\frac{w_{n_1}}{q_{n_1}}\bm{\theta}_{n_1}^t\|^2 = \frac{1}{K}\sum_{n=1}^{N}\frac{w_n^2}{q_n^2}q_n\|\bm{\theta}_n^t\|^2 \\
    &= \sum_{n=1}^{N}\frac{w_n^2}{Kq_n}\|\bm{\theta}_n^t\|^2,
\end{align*}
where~\eqref{in_exp_cauchy} follows from Cauchy-Schwarz inequality.
\end{IEEEproof}

%%%%%%%%%%%%%%%%%%%%%%%%%%%%%%%%%%%%%%%%
\begin{lemma}[Bounded Local Divergence]\label{lemma_boud_locl_divg} Under Assumptions \ref{ass:smoothness} and \ref{ass:bound_dissimi}, if the local learning rate $\eta\leq{1}/({2\sqrt{2}E \beta})$, then
the local model difference at round $t$ is bounded as follows:
\begin{align*}
    \sum_{n=1}^{N}\mathbb{E} [ w_n\|\bm{\theta}^t-\bm{\theta}_{n}^{t,s}\|^2] \leq 16\eta^2E^2\gamma^2\|\nabla F(\bm{\theta}^t)\|^2 + 16\eta^2E^2\kappa^2.
\end{align*}
\end{lemma}
%%%%%%%%%%%%%%%%%%%%%%%%%%%%%%%%%%%%%%%%%
\begin{IEEEproof} According to the local update rule, we have
\begin{align*}
    \mathbb{E}&\|\bm{\theta}^t-\bm{\theta}_{n}^{t,s}\|^2 = \mathbb{E}\|\bm{\theta}_{n}^{t,s-1}-\bm{\theta}^t-\eta\nabla F_n(\bm{\theta}_{n}^{t,s-1})\\&+\eta\nabla F_n(\bm{\theta}^{t})
    -\eta\nabla F_n(\bm{\theta}^{t})\|^2\\
    \labelrel\leq{in_boud_locl_divg_cahy} & (1+\frac{1}{2E-1})\mathbb{E}\|\bm{\theta}_{n}^{t,s-1}-\bm{\theta}^t\|^2\\
    &+2\eta^2E\mathbb{E}\|\nabla F_n(\bm{\theta}_{n}^{t,s-1})+\nabla F_n(\bm{\theta}^{t})-\nabla F_n(\bm{\theta}^{t})\|^2\\
    \labelrel\leq{in_boud_locl_divg_cahy1} & (1+\frac{1}{2E-1})\mathbb{E}\|\bm{\theta}_{n}^{t,s-1}-\bm{\theta}^t\|^2\\
    &+4\eta^2E\mathbb{E}\|\nabla F_n(\bm{\theta}_{n}^{t,s-1})-\nabla F_n(\bm{\theta}^{t})\|^2+4\eta^2E\mathbb{E}\|\nabla F_n(\bm{\theta}^{t})\|^2\\
    \labelrel\leq{in_boud_locl_divg_smoth} & (1+\frac{1}{2E-1})\mathbb{E}\|\bm{\theta}_{n}^{t,s-1}-\bm{\theta}^t\|^2\\
    &+4\eta^2\beta^2E\mathbb{E}\|\bm{\theta}_{n}^{t,s-1}-\bm{\theta}^{t}\|^2+4\eta^2E\mathbb{E}\|\nabla F_n(\bm{\theta}^{t})\|^2
\end{align*}
where~\eqref{in_boud_locl_divg_cahy} follows from Lemma~\ref{lemma_in_square_sum} with $\alpha={1}/({2E-1})$,~\eqref{in_boud_locl_divg_cahy1} follows from Cauchy-Schwarz inequality, and~\eqref{in_boud_locl_divg_smoth} follows from Assumption~\ref{ass:smoothness}. Next, we get
\begin{align*}
    &\sum_{n=1}^{N}\mathbb{E}w_n\|\bm{\theta}^t-\bm{\theta}_{n}^{t,s}\|^2 \labelrel\leq{in_boud_locl_divg_dsim} 4\eta^2E\gamma^2\mathbb{E}\|\nabla F(\bm{\theta}^{t})\|^2+4\eta^2E\kappa^2\\
    % & \leq  (1+\frac{1}{2E-1}+4\eta^2\beta^2E)\sum_{n=1}^{N}\mathbb{E} w_n\|\bm{\theta}_{n}^{t,s-1}-\bm{\theta}^t\|^2\\
    % &\quad+4\eta^2E\sum_{n=1}^{N}\mathbb{E} w_n\|\nabla F_n(\bm{\theta}^{t})\|^2\notag\\
    &   +(1+\frac{1}{2E-1}+4\eta^2\beta^2E)\sum_{n=1}^{N}\mathbb{E} w_n\|\bm{\theta}_{n}^{t,s-1}-\bm{\theta}^t\|^2
\end{align*}
where \eqref{in_boud_locl_divg_dsim} holds due to Assumption \ref{ass:bound_dissimi}. When $\eta\leq{1}/({2\sqrt{2}E \beta})$, then $4\eta^2\beta^2E\leq{1}/({2E})$, we have
\begin{align*}
    \sum_{n=1}^{N}\mathbb{E} w_n\|\bm{\theta}^t-\bm{\theta}_{n}^{t,s}\|^2 \leq & (1+\frac{1}{E-1})\sum_{n=1}^{N}\mathbb{E} w_n\|\bm{\theta}_{n}^{t,s-1}-\bm{\theta}^t\|^2\\
    &+4\eta^2E\gamma^2\mathbb{E}\|\nabla F(\bm{\theta}^{t})\|^2+4\eta^2E\kappa^2.
\end{align*}
Unrolling the recursion, we get
\begin{align*}
    &\sum_{n=1}^{N}\mathbb{E} w_n\|\bm{\theta}^t-\bm{\theta}_{n}^{t,s}\|^2 \\
    &\leq  \sum_{h=1}^{s-1}(1+\frac{1}{E-1})^h\left[4\eta^2E\gamma^2\mathbb{E}\|\nabla F(\bm{\theta}^{t})\|^2+4\eta^2E\kappa^2\right]\\
    &\leq  (E-1)\left[(1+\frac{1}{E-1})^E-1\right]\cdot[4\eta^2E\gamma^2\mathbb{E}\|\nabla F(\bm{\theta}^{t})\|^2\\
    &\quad+4\eta^2E\kappa^2]\\
    &\labelrel\leq{in_diff_mods_tau} 16\eta^2E^2\gamma^2\|\nabla F(\bm{\theta}^t)\|^2 + 16\eta^2E^2\kappa^2, 
\end{align*}
where \eqref{in_diff_mods_tau} follows from $(1+\frac{1}{E-1})^E\leq5$ when $E\geq1$.
\end{IEEEproof}

%---------------------------------
\subsection{Proof of Theorem~\ref{th:convergence}}
%---------------------------------
\begin{IEEEproof}
According to Assumption \ref{ass:smoothness}, we have
\begin{align}
    &\mathbb{E}  [F(\bm{\theta}^{t+1}-F(\bm{\theta}^t)] \notag\\
    &\leq   \mathbb{E}\langle \nabla F(\bm{\theta}^t),\bm{\theta}^{t+1}-\bm{\theta}^t\rangle+\frac{\beta}{2}\mathbb{E}\|\bm{\theta}^{t+1}-\bm{\theta}^t\|^2\notag\\
    & = \mathbb{E}\langle \nabla F(\bm{\theta}^t),\mathbb{E}_{\mathcal{K}^t}\sum_{n\in \mathcal{K}^t}\frac{w_n}{Kq_n}(\bm{\theta}_n^{t+1}-\bm{\theta}^t)\rangle\notag\\
    &\quad+\frac{\beta}{2}\mathbb{E}\|\sum_{n\in \mathcal{K}^t}\frac{w_n}{Kq_n}(\bm{\theta}_n^{t+1}-\bm{\theta}^t)\|^2\notag\\
    & \labelrel={eq_exp_samp_randk} \underbrace{-\mathbb{E}\langle \nabla F(\bm{\theta}^t),\sum_{n=1 }^{N}w_n (\eta\sum_{s=0}^{E-1}\nabla F_n(\bm{\theta}_{n}^{t,s}))\rangle}_{T_1}\notag\\
    &\quad+\underbrace{\frac{\beta}{2}\mathbb{E}\|\sum_{n\in \mathcal{K}^t}\frac{w_n}{Kq_n}\eta\sum_{s=0}^{E-1}\nabla F_n(\bm{\theta}_{n}^{t,s})\|^2}_{T_2},\label{eq_exp_samp}
\end{align}
where~\eqref{eq_exp_samp_randk} holds due to Lemma~\ref{lemma_exp_mom_sampling}.
For $T_1$, we have
\begin{align}
    T_1 
    % = & -\eta E\mathbb{E}\langle \nabla F(\bm{\theta}^t),\frac{1}{E}\sum_{n=1 }^{N} \sum_{s=0}^{E-1} w_n\nabla F_n(\bm{\theta}_n^{t,s})\rangle\notag\\
    \labelrel={t1_inner_prod} & -\frac{\eta E}{2}\|\nabla F(\bm{\theta}^t)\|^2-\frac{\eta E}{2}\|\frac{1}{E}\sum_{n=1 }^{N} \sum_{s=0}^{E-1} w_n\nabla F_n(\bm{\theta}_n^{t,s})\|^2\notag\\
    % &+\frac{\eta E}{2}\|\nabla F(\bm{\theta}^t)-\frac{1}{E}\sum_{n=1 }^{N} \sum_{s=0}^{E-1} w_n\nabla F_n(\bm{\theta}_n^{t,s})\|^2\notag\\
    % \leq & -\frac{\eta E}{2}\|\nabla F(\bm{\theta}^t)\|^2\notag\\
    % &+\frac{\eta E}{2}\|\nabla F(\bm{\theta}^t)-\frac{1}{E}\sum_{n=1 }^{N} \sum_{s=0}^{E-1} w_n\nabla F_n(\bm{\theta}_n^{t,s})\|^2 \notag\\
    % \labelrel={in_t1_jens} & -\frac{\eta E}{2}\|\nabla F(\bm{\theta}^t)\|^2\\
    &+\frac{\eta E}{2}\|\frac{1}{E}\sum_{n=1 }^{N} \sum_{s=0}^{E-1}w_n(\nabla F_n(\bm{\theta}^t)- \nabla F_n(\bm{\theta}_n^{t,s}))\|^2\notag\\
    \labelrel\leq{in_t1_jens} & -\frac{\eta E}{2}\|\nabla F(\bm{\theta}^t)\|^2\notag\\
    &+\frac{\eta }{2}\sum_{n=1 }^{N} \sum_{s=0}^{E-1}w_n\|\nabla F_n(\bm{\theta}^t)- \nabla F_n(\bm{\theta}_n^{t,s})\|^2\notag\\
    \labelrel\leq{in_t1_smooth} & -\frac{\eta E }{2}\|\nabla F(\bm{\theta}^t)\|^2+\frac{\eta \beta^2}{2}\sum_{n=1 }^{N} \sum_{s=0}^{E-1}w_n\|\bm{\theta}^t- \bm{\theta}_n^{t,s}\|^2,\label{t1}
\end{align}
where~\eqref{t1_inner_prod} follows from $2\langle \bm{a},\bm{b}\rangle = \|\bm{a}\|^2+\|\bm{b}\|^2-\|\bm{a}-\bm{b}\|^2$,~\eqref{in_t1_jens} holds due to Jensen's inequality, and~\eqref{in_t1_smooth} comes from Assumption~\ref{ass:smoothness}. For $T_2$, using Lemma~\ref{lemma_exp_mom_sampling}, we obtain:
\begin{align}
    T_2 
    % & \frac{\beta}{2}\mathbb{E}\|\sum_{n\in \mathcal{K}^t}\frac{w_n}{Kq_n}\eta\sum_{s=0}^{E-1}\nabla F_n(\bm{\theta}_{n}^{t,s})\|^2\notag\\
    = \frac{\beta}{2}\sum_{n=1}^{N}\frac{w_n^2\eta^2}{Kq_n}\|\sum_{s=0}^{E-1}\nabla F_n(\bm{\theta}_{n}^{t,s})\|^2 \leq \frac{\beta\eta^2E^2 G^2}{2K}\sum_{n=1}^{N}\frac{w_n^2}{q_n}.\label{t2_final}
\end{align}

Combining~\eqref{eq_exp_samp},~\eqref{t1} and~\eqref{t2_final}, we get:
\begin{align*}
     \mathbb{E}&  [F(\bm{\theta}^{t+1}-F(\bm{\theta}^t)] \leq  -\frac{\eta E }{2}\|\nabla F(\bm{\theta}^t)\|^2\\
     &+\frac{\eta \beta^2}{2}\sum_{n=1 }^{N} \sum_{s=0}^{E-1}w_n\|\bm{\theta}^t- \bm{\theta}_n^{t,s}\|^2 + \frac{\beta\eta^2E^2 G^2}{2K}\sum_{n=1}^{N}\frac{w_n^2}{q_n}\notag\\
     \leq &-\frac{\eta E }{2}\|\nabla F(\bm{\theta}^t)\|^2+ \frac{\beta\eta^2E^2 G^2}{2K}\sum_{n=1}^{N}\frac{w_n^2}{q_n}\notag\\
     &+\frac{\eta \beta^2E}{2}(16\eta^2E^2\gamma^2\|\nabla F(\bm{\theta}^t)\|^2 + 16\eta^2E^2\kappa^2)\\
     = & (-\frac{\eta E }{2}+8\eta^3\beta^2\gamma^2E^3)\|\nabla F(\bm{\theta}^t)\|^2+ 8\eta^3\beta^2E^3\kappa^2 \\
     &+ \frac{\beta\eta^2E^2 G^2}{2K}\sum_{n=1}^{N}\frac{w_n^2}{q_n}.\notag
\end{align*}
If taking $\eta \leq {1}/({32E^2 \beta^2\gamma^2})$, then we have:
\begin{align*}
     \mathbb{E}  [F(\bm{\theta}^{t+1}-F(\bm{\theta}^t)] \leq & -\frac{\eta E}{4}\|\nabla F(\bm{\theta}^t)\|^2 +  8\eta^3\beta^2E^3\kappa^2 \\
     &+ \frac{\beta\eta^2E^2 G^2}{2K}\sum_{n=1}^{N}\frac{w_n^2}{q_n}.\notag
\end{align*}
Summing the above inequality from $t=0,\dots,T-1$ and rearranging it, we obtain:
\begin{align}
    \sum_{t=0}^{T-1}\|\nabla F(\bm{\theta}^t)\|^2\leq & \frac{4(F(\bm{\theta}^{0})-F^*)}{\eta E}+8T\eta^2\beta^2E^2 \kappa^2 \\
    &+ \frac{2\beta\eta E G^2}{K}\sum_{t=0}^{T-1}\sum_{n=1}^{N}\frac{w_n^2}{q_n^t}. \notag
\end{align}
Dividing both sides by $T$, we get the final result.
% \begin{align}
% \frac{1}{T}\sum_{t=0}^{T-1}\|\nabla F(\bm{\theta}^t)\|^2\leq & \frac{4(F(\bm{\theta}^{0})-F^*)}{\eta TE}+8\eta^2\beta^2E^2 \kappa^2 + \frac{2\beta\eta E G^2}{KT}\sum_{t=0}^{T-1}\sum_{n=1}^{N}\frac{w_n^2}{q_n^t}\notag 
% \end{align}
\end{IEEEproof}

%------------------------------------------------
\section{Proof of Lemma~\ref{lemma_drift_plus_penalty}}\label{append_lemma1}
%------------------------------------------------
\begin{IEEEproof} By squaring the \textit{virtual} queue $Q_n^{t+1} = \max\{Q_n^{t} + a_n^t , 0\}$, we have:
\begin{align*}
(\max\{Q_n^{t} + a_n^t , 0\})^2 \leq (Q_n^t)^2 + (a_n^t)^2 + 2Q_n^ta_n^t.
\end{align*}
Then, we obtain:
\begin{align}\label{in_virtual_queue}
\frac{(Q_n^{t+1})^2 - (Q_n^{t})^2}{2} \leq \frac{(a_n^t)^2}{2} + Q_n^ta_n^t.
\end{align}
For $a_n^t$, we have:
\begin{align*}
a_n^t  &= (1-(1-q_{n}^{t})^{K}) (\mathcal{E}_{n}^{t,\text{cmp}} + \mathcal{E}_{n}^{t,\text{com}}) - \bar{\mathcal{E}}_{n}\\
& = (1-(1-q_{n}^{t})^{K}) ( \frac{E\alpha_{n}c_nD_n(f_n^t)^2}{2} + p_{n}^{t}\mathcal{T}_{n,u}^{t,\text{com}}) - \bar{\mathcal{E}}_{n}.
\end{align*} 
According to the formula $(x-y)^2 \leq x^2+y^2 (x,y \geq 0)$ and $(1-(1-q_{n}^{t})^{K}) \leq 1$, we have:
\begin{align*}
(a_n^t)^2 &\leq \left[ \frac{E\alpha_{n}c_nD_n(f_n^t)^2}{2} + p_{n}^{t}\mathcal{T}_{n,u}^{t,\text{com}}\right]^2 + (\bar{\mathcal{E}}_{n})^2\\
&\leq \left[ \bar{\mathcal{T}}_n^{t,\text{com}} p_{n}^{\max} +  \frac{E\alpha_{n}c_nD_n(f_n^{\max})^2}{2}\right]^2 + (\bar{\mathcal{E}}_{n})^2
\end{align*}
where $\bar{\mathcal{T}}_n^{t,\text{com}}$ is the upper bound of $\mathcal{T}_{n,u}^{t,\text{com}}$. Summing over all devices, taking the expectation w.r.t $\bm{Q}^t$ on both sides, and adding penalty term $V\mathbb{E}\left\{\sum_{n=1}^{N} \left(q_{n}^{t}\mathcal{T}_{n}^{t}+\lambda \cdot  \frac{w_n^2}{ q_{n}^{t}}\right)|\bm{Q}^t\right\}$ to both sides of the inequality~\eqref{in_virtual_queue}, we get the final result.
\end{IEEEproof}

%------------------------------------------------------
\section{Proof of Theorem~\ref{theo_solu_fnt}} \label{append_theo_solu_fnt}
%------------------------------------------------------
\begin{IEEEproof} Problem \textbf{3.1.1} can be rewritten as
\begin{align}
&\min_{\{f_n^t\}}\quad \sum_{n=1}^{N}\frac{\Omega_{1,n}}{f_{n}^{t}} + \sum_{n=1}^{N}\Omega_{2,n}(f_{n}^{t})^{2} \notag\\
& \text{s.t.}  \qquad f_{n}^{\min} \leq f_{n}^{t} \leq f_{n}^{\max},\quad \forall n\in[N], \notag
\end{align}
where $\Omega_{1,n}$ and $\Omega_{2,n}$ are positive constants denoted by $\Omega_{1,n} = VEq_n^tc_nD_n, \Omega_{2,n} = \frac{1}{2}Q_n^t (1-(1-q_{n}^{t})^{K})E\alpha_{n}c_{n}D_n$. Taking the first-order derivation of the objective function, we obtain
\begin{align}\label{eq_root_fnt_a12_1}
    -\frac{\Omega_{1,n}}{(f_n^t)^2} + 2\Omega_{2,n}f_n^t = 0,\quad
    f_n^t = \left(\frac{\Omega_{1,n}}{2\Omega_{2,n}}\right)^{\frac{1}{3}}
\end{align}
Then, for the second-order derivation, we have
\begin{align*}
    2\frac{\Omega_{1,n}}{(f_n^t)^3} + 2\Omega_{2,n} > 0.\notag
\end{align*}
So the root in~\eqref{eq_root_fnt_a12_1} is the global minimum. In final, we obtain
\begin{align*}
    (f_n^t)^* = \min\{ \max\{\left(\frac{Vq_n^t}{Q_n^t \big[(1-(1-q_{n}^{t})^{K}]\alpha_{n}}\right)^{\frac{1}{3}}, f_{n}^{\min}\} , f_{n}^{\max} \}.
\end{align*}
\end{IEEEproof}

%------------------------------------------------------
\section{Proof of Theorem~\ref{theo_solu_punt}} \label{append_theo_solu_pnt}
%------------------------------------------------------

% \begin{align}\label{prob_pnt_appen}
% &\min_{\{p_n^t\}} \quad
% \sum_{n=1}^{N} \frac{MK}{B \log_2(1+\frac{h_{n}^{t}p_{n}^{t}}{N_{0}})} \big(Q_n^t (1-(1-q_{n}^{t})^{K}) p_{n}^{t} + Vq_{n}^{t}\big)\\
% & \quad \text{s.t.} \quad\qquad\quad  p_{n}^{\min} \leq p_{n}^{t} \leq p_{n}^{\max},\quad \forall n\in[N] \notag
% \end{align}
\begin{IEEEproof}
Problem \textbf{P3.1.2} can be rewritten as
\begin{align}\label{prob_pnt_An}
\min_{\{x_n^t\}}\quad &\sum_{n=1}^{N}\Omega_{3,n}\frac{x_n^t+A_{1,n}}{\log_{2}{(1+x_n^t})} \\
 \text{s.t.}  \quad &\frac{h_n^tp_n^{\min}}{N_0} \leq x_{n}^{t} \leq \frac{h_n^tp_n^{\max}}{N_0},\quad \forall n\in[N], \notag
\end{align}
where $x_n^t = \frac{h_n^tp_n^t}{N_0}$, $\Omega_{3,n} = \frac{MKN_0}{Bh_n^t}Q_{n}^{t}(1-(1-q_n^t)^K)$ and $A_{1,n}= \frac{Vq_n^th_n^t}{Q_{n}^{t}(1-(1-q_n^t)^K)N_0}$ are always positive. Problem~\eqref{prob_pnt_An} can be solved by individually optimizing $x_n^t$. So taking the second-order derivation of the objective~\eqref{prob_pnt_An}, we obtain
\begin{align}
    \Omega_{3,n}\frac{(A_{1,n}-x_n^t-2)\ln(1+x_n^t)+2(A_{1,n}+x_n^t)}{(1+x_n^t)^2\ln^2{(2)}\log_2^3{(1+x_n^t)}}
\end{align}
From the inequality $\ln(1+x) \geq \frac{x}{1+x} $, ($x>=0$) we have:
\begin{align}
    &(A_{1,n}-x_n^t-2)\ln(1+x_n^t)+2(A_{1,n}+x_n^t)\notag\\
    &\geq \frac{(A_{1,n}-x_n^t-2)x_n^t+2(A_{1,n}+x_n^t)(1+x_n^t)}{1+x_n^t}\label{pnt_2nd_low}
\end{align}
Simplify the numerator of~\eqref{pnt_2nd_low}. We obtain
\begin{align}
    (x_n^t)^2 + 3A_{1,n}x_n^t + 2A_{1,n}\geq 0,
\end{align}
where the inequality holds because $x_n^t, A_{1,n}$ are positive. Thus, Problem~\eqref{prob_pnt_An} is convex. To solve the problem, we take the first order derivation of the objective function~\eqref{prob_pnt_An}, we have
\begin{align}\label{eq_xnt_solu}
     \Omega_{3,n}\frac{\log_2(1+x_n^t)-\frac{(x_n^t+A_{1,n})}{\ln{2}(1+x_n^t)}}{\log_2^2{(1+x_n^t)^2}}=0
\end{align}
\begin{align*}
    \ln{(1+\frac{h_n^tp_n^t}{N_0})}=\frac{(\frac{h_n^tp_n^t}{N_0}+A_{1,n})}{(1+\frac{h_n^tp_n^t}{N_0})}
\end{align*}
By solving \eqref{eq_xnt_solu} and definition of $x_n^t$, we get
\begin{align}\label{eq_root_pnt_a5}
    \ln{(1+\frac{h_n^tp_n^t}{N_0})} = \frac{h_n^tp_n^t + A_{1,n}N_0}{h_n^tp_n^t+N_0}
\end{align}
So the root in~\eqref{eq_root_pnt_a5} is global minimum in the first quadrant. In final, we obtain
\begin{align*}
    (p_n^t)^* = \min\{ \max\{(p_n^t)^\prime, p_{n}^{\min}\} , p_{n}^{\max} \},
\end{align*}
where $(p_n^t)^\prime$ is the root of~\eqref{eq_root_pnt_a5}.
\end{IEEEproof}

\end{document}